\documentclass[12pt,draftclsnofoot, onecolumn]{IEEEtran}
\usepackage{amssymb}
\usepackage[cmex10]{amsmath}
\usepackage{stfloats}
\usepackage{graphicx}
\usepackage{subfigure}
\usepackage{tabularx}
\usepackage{epsfig,epsf,color,balance,cite}
\usepackage{verbatim}
\usepackage{url}
\usepackage{bm}
\usepackage{epstopdf}

\newtheorem{theorem}{\bf Theorem}

\newtheorem{lemma}{\bf Lemma}

\newtheorem{definition}{\bf Definition}

\usepackage{algorithm}
\usepackage{algorithmic}

\usepackage{color}
\definecolor{myc1}{rgb}{0,0,0}

\definecolor{myc2}{rgb}{0,0,0}
\begin{document}

\title{Beamforming Design for Multiuser Transmission Through Reconfigurable Intelligent Surface}

\author{
\IEEEauthorblockN{Zhaohui Yang,
                  Wei Xu, \IEEEmembership{Senior Member, IEEE},
Chongwen Huang,
Jianfeng Shi, and
                  Mohammad Shikh-Bahaei, \IEEEmembership{Senior Member, IEEE}
                  \vspace{-3em}}
\thanks{Z. Yang  and M. Shikh-Bahaei are with the Centre for Telecommunications Research, Department of Engineering, King's College London, WC2R 2LS, UK. (Emails: yang.zhaohui@kcl.ac.uk, m.sbahaei@kcl.ac.uk)}
\thanks{ W. Xu is with the National Mobile Communications Research
Laboratory, Southeast University, Nanjing 210096, China.  (Email: wxu@seu.edu.cn.)}
\thanks{C. Huang is with the Singapore University of Technology and Design, 487372 Singapore. (Email: chongwen\_huang@alumni.sutd.edu.sg)}
\thanks{J. Shi is with School of Electronic and Information Engineering, Nanjing University of Information Science and Technology, Nanjing 210096, China. (Email: jianfeng.shi@nuist.edu.cn)}
 }

\maketitle

\begin{abstract}
This paper investigates the problem of resource allocation for multiuser communication networks with a reconfigurable intelligent surface (RIS)-assisted wireless transmitter. In this network, the sum transmit power of the network is minimized by controlling the phase beamforming of the RIS and transmit power of the BS. This problem is posed as a joint optimization problem of transmit power and RIS control, whose goal is to minimize the sum transmit power under signal-to-interference-plus-noise ratio (SINR) constraints of the users. To solve this problem, a dual method is proposed, where the dual problem is obtained as a semidefinite programming problem. After solving the dual problem, the  phase beamforming of the RIS is obtained in the closed form, while the optimal transmit power is obtained by using the standard interference function. Simulation results show that the proposed scheme can reduce up to 94\%  and 27\%  sum transmit power compared to the maximum ratio transmission (MRT) beamforming and zero-forcing (ZF) beamforming techniques, respectively.
\end{abstract}

\IEEEpeerreviewmaketitle

\section{Introduction}

Driven by the rapid development of advanced multimedia applications, it is urgent for the next-generation wireless network to support high spectral efficiency and massive connectivity~\cite{saad2019vision}.
Due to the demand of high data rate and serving a massive number of users, energy consumption has become a challenging problem in the design of the future wireless network \cite{yu2007transmitter,ngo2013energy,buzzi2016survey,zhang2019first,7264975}.

Recently, reconfigurable intelligent surface (RIS)-assisted wireless communication has been proposed as a potential solution for enhancing the energy efficiency of wireless networks \cite{huang2020thzris,pan2019intelligent2,pan2019intelligent,liu2020reconfigurable,chongwenDL2020,huang2019holographic,yang2020energyefficient,
di2020smart,yang2020RSMARIS,long2020reflections}. 
RIS is a new paradigm
that can flexibly manipulate electromagnetic (EM) waves.
Researches of RIS-assisted wireless communications mainly follow into two aspects: RIS as a passive reflector and RIS as an active
transceiver.

On one hand, RIS can serve as a passive reflector.
An RIS is a meta-surface equipped with low-cost and passive elements that can be programmed to turn wireless channels into a partially deterministic space.
In RIS-assisted wireless communication systems, a base station (BS) sends control signals to an RIS controller so as to optimize the properties of incident waves and improve the communication quality of users \cite{zhou2020spectral,hu2020programmable,zhang2020sum}.
The RIS acting as a reflector does not perform any decoding or digitalization operation.
Hence, if properly deployed, the RIS promises much lower energy consumption than traditional amplify-and-forward (AF) relays~\cite{hum2013reconfigurable,huang2014relay,ntontin2019reconfigurable,jung2019optimality,
yu2020power,xu2020resource,8855810,hua2020intelligent}.
A number of existing works such as in \cite{huang2018achievable,jung2018performance,pan2019intelligent,zhao2019intelligent,8743496,yu2019robust,
9133435,xu2019resource,9133120,guan2020joint} have studied to optimize the deployment of RISs in wireless networks.
In \cite{huang2018achievable},
the downlink sum-rate of an RIS-assisted wireless communication system was characterized.
Asymptotic analysis of uplink data rate in an
RIS-based large antenna-array system was presented in \cite{jung2018performance}.
Considering energy harvesting, an RIS was invoked for enhancing the sum-rate performance of a system with simultaneous wireless information and power transfer \cite{pan2019intelligent}.
Instead of considering the availability of instantaneous channel state information (CSI), the authors in \cite{zhao2019intelligent} proposed  a two-time-scale transmission protocol to maximize the average achievable sum-rate for an RIS-aided multiuser system under a general correlated Rician channel model.
Taking the secrecy into consideration, the work in \cite{8743496} investigated the problem of secrecy rate maximization of an RIS-assisted multi-antenna system.
Further considering imperfect CSI, the physical layer security was enhanced by an RIS in a wireless channel~\cite{yu2019robust}.
Beyond the above studies,
the use of RISs for enhanced wireless energy efficiency has been studied in \cite{8741198}.
In \cite{8741198}, authors proposed a new approach to maximize the energy efficiency of a multiuser multiple-input single-output (MISO) system by jointly controlling the transmit power of the BS and the phase shifts of the RIS.
{\color{myc1}
The RIS-assisted simultaneous wireless information and power transfer (SWIPT) system was studied in \cite{9133435}, where the sum transmit power at the BS was minimized via jointly optimizing its transmit precoders and the reflect phase shifts at all RISs, subject to the quality-of-service (QoS) constraints at all users.
The authors in \cite{xu2019resource} studied the resource allocation design for secure communication in RIS-assisted multiuser  MISO communication systems by using  artificial noise (AN).
Considering both security and SWIPT, the energy efficiency maximization problem was studied in \cite{9133120} for the secure RIS-aided SWIPT.
For spectrum sensing, an RIS-assisted cognitive radio system was investigated in \cite{guan2020joint}, where an RIS is deployed to assist in the spectrum sharing between a primary user link and a secondary user link.}

On the other hand, the RIS-assisted wireless transmitter \cite{zhao2018programmable,dai2019wireless,tang2019programmable,basar2019transmission,basar2019wireless,tang2019mimo} can directly perform
modulation on the EM carrier signals, without the need for conventional radio-frequency (RF) chains, which can be used for holographic multiuser multiple-input multiple-output (MIMO) technologies.
In \cite{dai2019wireless}, authors investigated the RIS-based quadrature phase shift keying (QPSK) transmission over wireless channels.
The RIS-based
8-phase shift keying (8PSK) was further studied in \cite{tang2019programmable}.
The feasibility of using RIS for MIMO with higher-order modulations was studied in  \cite{tang2019mimo}, which presented an analytical modelling of the RIS-based system.
However, the above works  \cite{zhao2018programmable,dai2019wireless,tang2019programmable,basar2019transmission,basar2019wireless,tang2019mimo} only  considered the RIS-assisted wireless transmitter for single-user cases.
In this paper, we investigate the beamforming design for multiuser transmission with the RIS-assisted wireless transmitter.
The main contributions of this paper include:
\begin{itemize}
\item
 We consider a downlink wireless communication system with one RIS-assisted wireless transmitter and multiple users.
 To minimize the sum transmit power of the BS,
we jointly optimize phase shifts of the RIS and the multiuser power allocation at the BS.
  We formulate an optimization
problem with the objective of minimizing the sum transmit power under individual user constraints in terms of signal-to-interference-plus-noise ratio (SINR) and  unit-modulus constraint of the RIS phase shifts.
\item To minimize the sum transmit power of the BS, a dual method is proposed.
By using the dual method, the dual problem of the sum transmit power minimization problem is first obtained.
Then, phase shifts of the RIS can be obtained in the closed form. For the transmit power of the BS, an iterative power control scheme based on the standard interference function is proposed to obtain the optimal power control.
\item We consider both maximum ratio transmission (MRT) beamforming and zero-forcing (ZF) beamforming techniques when solving the sum transmit power minimization problem. The power scaling law performance of the multiuser communication is evaluated for the RIS-assisted wireless transmitter. Simulation results show that the proposed method saves up to 94\%  and
27\% sum transmit power compared to the conventional MRT and ZF schemes, respectively.
\end{itemize}

The rest of this paper is organized as follows.
System model and problem formulation are described in Section \uppercase\expandafter{\romannumeral2}.
Section \uppercase\expandafter{\romannumeral3} provides
the  algorithm design.
Simulation results are presented in Section \uppercase\expandafter{\romannumeral4}. Conclusions are drawn in Section \uppercase\expandafter{\romannumeral5}.

Notations: In this paper, the imaginary unit of a complex number is denoted by $j=\sqrt{-1}$.
Matrices and vectors are denoted by boldface capital and lower-case letters, respectively.
Matrix $\text{diag}(x_1,\cdots,x_N)$ denotes a diagonal matrix whose diagonal components
are $x_1,\cdots,x_N$.
The real part of a complex number $x$ is denoted by $\mathcal R(x)$.
$\boldsymbol X \succeq \boldsymbol 0$ indicates that
$\boldsymbol X$ is a positive semidefinite matrix.
$\boldsymbol x^*$, $\boldsymbol x^T$, and $\boldsymbol x^H$ respectively denote the conjugate, transpose, and
conjugate transpose of vector $\boldsymbol x$.
$[\boldsymbol x]_{n}$ and $[\boldsymbol X]_{kn}$ denote the $n$-th and $(k,n)$-th elements of the
respective vector $\boldsymbol x$ and matrix $\boldsymbol X$.
$|x|$ stands for the module of a complex number $x$, while $\|\boldsymbol x\|$ denotes the $\ell_2$-norm of vector $\boldsymbol x$.
The identity matrix is denoted by $\boldsymbol I$, while an all-one vector is denoted by $\boldsymbol 1$.
The distribution of a circularly symmetric complex
Gaussian variable with mean $x$ and covariance $\sigma$ is denoted by ${\mathcal {CN}}( x, \sigma)$.
The expectation operation is denoted by $\mathbb E$.
$\boldsymbol X^\dagger$ denotes the Moore-Penrose pseudoinverse of matrix $\boldsymbol X$.
The optimal value of an optimization variable $\boldsymbol X$ is denoted by $\boldsymbol X^\star$.
\section{System Model and Problem Formulation}

Consider an RIS-assisted multiuser wireless communication system that consists of {\color{myc1}one BS with a single antenna}, an RIS, and a set $\mathcal K$ of $K$ users.
The RIS consists of $M$ rows and $N$ columns of RIS units, while
each user is equipped with one antenna.
The RIS unit in the $k$-th row and $n$-th column is denoted by $U_{kn}$.

\subsection{RIS-Assisted Modulation}
\begin{figure}[t]
\centering
\includegraphics[width=5in]{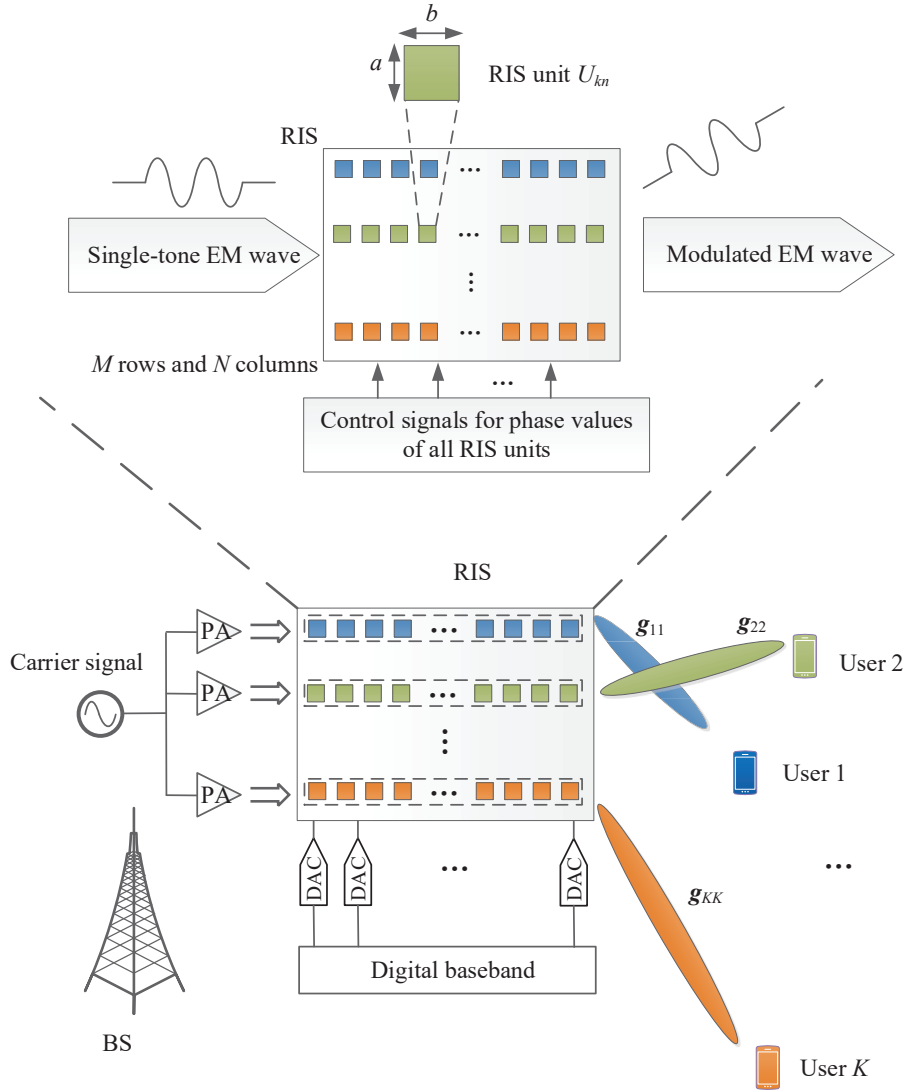}
\caption{{\color{myc1}An RIS-assisted multiuser wireless communication system.}} \label{figsys1}
\end{figure}

RIS is indeed a programmable metasurface composed of sub-wavelength
units, which can manipulate EM waves.
All RIS units are regularly
arranged  in a two-dimensional structure, as shown in Fig. \ref{figsys1}.
The length and width of each RIS unit are denoted by $a$ and $b$, respectively.
The control signal of each RIS unit can change the electrical parameter of
the tunable component, such as the phase \cite{tang2019mimo}.

Let $\bar E$ and $\hat E$ denote the incident and reflecting EM waves on RIS unit $U_{kn}$, respectively.
Following the definition of reflection phase of an EM, say $\phi_{kn}$ as the reflection phase of RIS unit $U_{kn}$, it follows:
\begin{equation}\label{sys1eq0}
\hat E =\text e^{j \phi_{kn}} \bar E.
\end{equation}
In particular, if the incident EM wave $\bar E$ is a single-tone EM wave (i.e., a carrier signal) with frequency $f_{\text c}$ and amplitude $A_{\text c}$, equation \eqref{sys1eq0} can be further expressed by:
\begin{equation}\label{sys1eq0_1}
\hat E =A_{\text c}\text e^{j (2\pi f_{\text c} t+\phi_{kn})}.
\end{equation}
From \eqref{sys1eq0_1}, it can be observed that the adjustable phase $\phi_{kn}$  can achieve a phase
modulation on the carrier signal, which is referred to as the RIS-assisted modulation.

\subsection{RIS-Assisted Multiuser Communication}
An RIS-assisted multiuser wireless communication system\footnote{{\color{myc1}In system model, the authors in \cite{wu2019intelligent} considered RIS as a passive relay, while RIS is used as a passive transmitter in this paper.
Due to this difference, the mathematical signal model in this paper is different from \cite{wu2019intelligent}.
}} is illustrated in Fig. \ref{figsys1} \cite{dai2019wireless,tang2019mimo}.
{\color{myc1}The reflection phase of each RIS unit is controlled by the digital baseband through the digital-to-analog converters (DACs), i.e., each RIS unit is controlled by one dedicated DAC. Before the carrier signal is reflected to each row of the RIS units, one narrow-band power amplifier (PA) is used to control the power of the carrier signal and the energy flux density on each RIS unit in the $k$-th row is denoted as $D_k$. The number of PAs equal to the number of served users.
Note that in the considered system, the information is actually
embedded in the phase control signals of each row of the RIS.
As shown in Fig. \ref{figsys1}, RIS is not connected with the digital baseband directly and RIS is used to reflect the baseband signal.

Compared  to the traditional hybrid mmWave system, the differences of the current RIS-transmitter setup include:
\begin{enumerate}
  \item Transmit antenna: There are multiple transmit antennas in  the traditional hybrid mmWave system, while only one transmit antenna is needed for the RIS-assisted wireless transmitter system in this paper.
  \item Modulation: For the traditional hybrid mmWave system, the signals are modulated in multiple antennas before the baseband precoder. The RIS-assisted wireless transmitter can directly perform modulation on the EM carrier signals, i.e., continuous phase modulation scheme is adopted.
  \item RF chain: For the traditional hybrid mmWave system, multiple RF chains are needed. In contrast, the proposed RIS-assisted wireless transmitter is considered as an RF chain-free transmitter.
\end{enumerate}
}

To construct the multiple data streams for multiple users, the $N$ units in the $k$-th row of the RIS are allocated to user $k$\footnote{In this paper, the RIS unit allocation for each user is assumed to be fixed for the convenience of deployment, i.e., $K=M$.}, i.e., the phases of RIS units $U_{k1}, U_{k2}, \cdots, U_{kN}$ are used to modulate the transmitted signal for user $k$.
Let the transmitted signal $s_k$ for user $k$ be:
\begin{equation}\label{sys1eq0_2}
s_k=\text e^{j \varphi_{k}}.
\end{equation}
Through changing the value of $\varphi_k$, signal $s_k$ can be modulated by  phase shift-keying (PSK).
In \cite{tang2019mimo}, it was shown that quadrature amplitude modulation (QAM) can also by achieved by equation \eqref{sys1eq0_2} of proper designs.
As a result, the transmitted signal by the RIS units $U_{k1}, U_{k2}, \cdots, U_{kN}$ can be presented by:
\begin{align}\label{sys1eq0_3}
[\text e^{j \phi_{k1}},\text e^{j \phi_{k2}},\cdots,\text e^{j \phi_{kN}}]^T
&=
[\text e^{j \varphi_{k1}},\text e^{j \varphi_{k2}},\cdots,\text e^{j \varphi_{kN}}]^T\text e^{j \varphi_{k}}
\nonumber\\&= [\text e^{j \varphi_{k1}},\text e^{j \varphi_{k2}},\cdots,\text e^{j \varphi_{kN}}]^T s_k,
\end{align}
where $\varphi_{kn}=\phi_{kn}-\varphi_k$.
For notational simplicity, we introduce:
\begin{equation}\label{sys1eq0_5}
\theta_{kn}\triangleq\text {e}^{j\varphi_{kn}},
\end{equation}
and equation \eqref{sys1eq0_3} can be rewritten as
\begin{align}
[\text e^{j \varphi_{k1}},\text e^{j \varphi_{k2}},\cdots,\text e^{j \varphi_{kN}}]^T s_k
=\boldsymbol \theta_k s_k,
\end{align}
where $\boldsymbol \theta_k=[\theta_{k1},\cdots, \theta_{kN}]^T$ is the phase beamformer of user $k$ which can be adjusted by the RIS.

Assume that the carrier signal with frequency $f_{\text c}$ is a uniform plane wave.
In consequence, the transmitted  signal at the BS is
\begin{equation}\label{sys1eq1}
\boldsymbol x=[ \sqrt{p_1}s_1\boldsymbol \theta_1;\sqrt{p_2}s_2\boldsymbol \theta_2;\cdots;\sqrt{p_K}s_K\boldsymbol \theta_K],
\end{equation}
where $p_k=(abD_k)^2$ is the transmit power of the BS for user $k$.
{\color{myc1}Equation \eqref{sys1eq1} is a mathematical signal model.}
%
%




{\color{myc1}Since the source is close to the RIS, as shown in Fig. \ref{figsys1},  the channel from the source to the RIS can be precisely measured, which can be regarded as a constant as in \cite{tang2019mimo}.}
The received signal at user $k$ can be given by:
\begin{equation}\label{sys1eq2}
y_k = \boldsymbol g_{k}^H \boldsymbol x+n_k=\sum_{i=1}^K \sqrt{p_i}\boldsymbol g_{ki}^H\boldsymbol \theta_i s_i +n_k,
\end{equation}
where $\boldsymbol g_{k}$ is the channel gain from all RIS elements to user $k$, $\boldsymbol g_{ki}$ is the channel gain from $N$ RIS elements in the $i$-th row to user $k$, and  $n_k\sim\mathcal {CN}(0,\sigma^2)$ is the additive white Gaussian noise.
{\color{myc1}Assume that the signals are only reflected by the RIS once.
Since both the phase and transmit power can be changed, we can construct that the input term  $\sqrt{p_i} s_i$ follows the Gaussian distribution, and the channel capacity can be achieved.}
Based on  this consideration on \eqref{sys1eq2},
the SINR at user $k$ is
\begin{equation}\label{sys2eq3_1}
\gamma_k=\frac{p_k \left| \boldsymbol g_{kk}^H\boldsymbol \theta_k \right|^2}
{\sum_{i=1,i\neq k}^K  p_i \left|\boldsymbol g_{ki}^H\boldsymbol \theta_i  \right|^2+\sigma^2}.
\end{equation}

{\color{myc1}Note that the required CSI for the proposed RIS-assisted wireless transmitter only includes channel gains from the BS (i.e., RIS) to users. In contrast, for the conventional RIS-assisted wireless transmission where the RIS acts as a passive reflector, the required CSI includes channel gains from the BS to the RIS, the RIS to users, and the BS to users.
As a result, compared to the conventional RIS-assisted wireless transmission, one key novelty of the proposed RIS-assisted wireless transmitter is that the required amount of CSI is smaller.
There are already many significant
methods that are proposed in existing works for obtaining CSI in RIS-based communication systems. For example, the works \cite{taha2019enabling} and \cite{huang2019indoor} presented compressive sensing and deep learning approaches for recovering the involved channels and designing the RIS phase matrix.
Based on the parallel factor framework,
the authors in \cite{wei2020parallel} proposed an alternating least
square method, which continuously estimates the
all channels without too high complexity.}

\subsection{Problem Formulation}
Given the considered system model,  our objective is to jointly optimize the phase beamforming $\boldsymbol \theta_k$ and transmit  power $p_k$ so as to minimize the sum transmit power
under individual minimum SINR requirements. Mathematically, the problem for the RIS-assisted multiuser transmission can be formulated as:
\begin{subequations}\label{sys2max1}
\begin{align}
\mathop{\min}_{ \boldsymbol\theta,    \boldsymbol p} \:&
 \sum_{k=1}^K  p_k \tag{\theequation}\\
\textrm{s.t.} \:
& \frac{p_k \left| \boldsymbol g_{kk}^H\boldsymbol \theta_k \right|^2}
{\sum_{i=1,i\neq k}^K  p_i \left|\boldsymbol g_{ki}^H\boldsymbol \theta_i  \right|^2+\sigma^2}\geq \Gamma_k, \quad \forall k\in\mathcal K,\\
&|\theta_{kn}| =1,   \quad  \forall k\in\mathcal K, n\in\mathcal N,
\end{align}
\end{subequations}
where $\boldsymbol\theta=[\theta_{11},\cdots,\theta_{1N},\cdots, \theta_{KN}]^T$, $\boldsymbol p=[p_1,\cdots, p_K]^T$,
$\mathcal N=\{1,\cdots,N\}$,
 and $\Gamma_k$ is the minimum SINR requirement of user $k$.
The minimum SINR constraints for all users are given in (\ref{sys2max1}a), and (\ref{sys2max1}b) presents the unit-modulus constraints.
Different from the conventional RF beamforming design, the phase beamforming problem (\ref{sys2max1}) introduces the unique unit-modulus constraints (\ref{sys2max1}b).
Due to the nonconvex constraints in (\ref{sys2max1}b), the problem  (\ref{sys2max1}) is nonconvex.

\section{Algorithm Design}
To solve the nonconvex  problem in  (\ref{sys2max1}), the dual method is first applied, where the dual problem of (\ref{sys2max1}) is always a convex problem, which can be effectively solved.
For comparisons, two conventional techniques, MRT beamforming and ZF beamforming, are also provided to solve the problem  (\ref{sys2max1}).

%

\subsection{Dual Method}
To rewrite sum power minimization problem  (\ref{sys2max1}) in a simplifier manner, we introduce $\boldsymbol w_k=\sqrt{p_k}\boldsymbol\theta_k$.
Replacing $\boldsymbol \theta_k$ with $\boldsymbol w_k$, the problem  (\ref{sys2max1}) is equivalent to:
\begin{subequations}\label{Alo3max1}
\begin{align}
\mathop{\min}_{ \boldsymbol w,    \boldsymbol p} \:&
 \sum_{k=1}^K  p_k \tag{\theequation}\\
\textrm{s.t.} \:
& \frac{ \left| \boldsymbol g_{kk}^H\boldsymbol w_k \right|^2}
{\sum_{i=1,i\neq k}^K   \left|\boldsymbol g_{ki}^H\boldsymbol w_i  \right|^2+\sigma^2}\geq \Gamma_k, \quad \forall k\in\mathcal K,\\
&  [\boldsymbol w_k\boldsymbol w_k^H]_{nn}=p_k,   \quad  \forall k\in\mathcal K, n\in\mathcal N,
\end{align}
\end{subequations}
where $\boldsymbol w=[w_{11},\cdots,w_{1N},\cdots, w_{KN}]^T$.

{\color{myc1}Denote $\boldsymbol \Gamma=[\Gamma_{1},\cdots, \Gamma_{K}]^T$ and we define the
following time-sharing condition.
\begin{definition}
Let $(\boldsymbol w^{(a)},    \boldsymbol p^{(a)})$ and $(\boldsymbol w^{(b)},    \boldsymbol p^{(b)})$ be the optimal solutions to the optimization problem \eqref{Alo3max1} with $\boldsymbol \Gamma=\boldsymbol \Gamma^{(a)}$ and $\boldsymbol \Gamma=\boldsymbol \Gamma^{(b)}$.
An optimization problem of the form \eqref{Alo3max1} is said to satisfy
the time-sharing condition if for any $\boldsymbol \Gamma^{(a)},\boldsymbol \Gamma^{(b)}$ and any $\kappa\in[0,1]$, there always exists a feasible solution $(\boldsymbol w^{(c)},    \boldsymbol p^{(c)})$ such that
\begin{equation*}
\frac{ \left| \boldsymbol g_{kk}^H\boldsymbol w_k^{(c)} \right|^2}
{\sum_{i=1,i\neq k}^K   \left|\boldsymbol g_{ki}^H\boldsymbol w_i^{(c)}  \right|^2+\sigma^2}\geq \kappa\Gamma_k^{(a)}+(1-\kappa)\Gamma_k^{(b)}, \quad \forall k\in\mathcal K,
\end{equation*}
\begin{equation*}
\boldsymbol w_k^{(c)}(\boldsymbol w_k^{(c)})^H]_{nn}=p_k^{(c)},   \quad  \forall k\in\mathcal K, n\in\mathcal N,
\end{equation*}
and
\begin{equation*}
 \sum_{k=1}^K  p_k ^{(c)} \geq \kappa  \sum_{k=1}^K  p_k ^{(a)} +(1-\kappa) \sum_{k=1}^K  p_k ^{(b)}.
\end{equation*}
\end{definition}
The time-sharing condition has the following intuitive interpretation.
Consider the maximum value of the optimization
problem \eqref{Alo3max1} as a function of the constraint $\boldsymbol \Gamma$. Clearly, a smaller $\Gamma_k$
implies a more relaxed constraint. So, roughly speaking,
the maximum value is an increasing function of $\boldsymbol \Gamma$. The
time-sharing condition implies that the maximum value of the
optimization problem is a concave function of $\boldsymbol \Gamma$.}
To show the gap between  problem \eqref{Alo3max1} and its dual problem, we provide the following lemma.
\begin{lemma}
The duality gap for multiuser RIS-assisted optimization \eqref{Alo3max1}
always tends to zero as the number of users $K$ or the number of RIS unit elements $N$ goes
to infinity, regardless of whether the original problem is
convex.
\end{lemma}
\itshape {Proof:}  \upshape
This result holds by directly applying Theorems 1 and 2 in \cite{1658226}.
According to \cite[Theorem~2]{1658226}, the time-sharing property holds for all optimization
problems with infinite channel gains, i.e., $K\rightarrow \infty$ or $N\rightarrow \infty$.
Based on \cite[Theorem~1]{1658226},   the optimization problem has a zero duality gap If it satisfies the time-sharing property.
\hfill $\Box$

According to Lemma 1, the near optimal solution of problem \eqref{Alo3max1} can be obtained by solving its dual problem if the number of users is high or the number of RIS units is large.
In practice, for example, since RIS unit is low-cost, a large number of RIS units can be deployed at the BS.
Consequently, the sum power minimization problem \eqref{Alo3max1} can be effectively solved via its dual problem.


\begin{theorem}
The dual problem of problem \eqref{Alo3max1} is:
\begin{subequations}\label{Alo3max2}
\begin{align}
\mathop{\max}_{ \boldsymbol q,    \boldsymbol \alpha} \:&
 \sum_{k=1}^K  \alpha_k \sigma^2 \tag{\theequation}\\
\textrm{s.t.} \:
& \sum_{n=1}^Nq_{kn}\leq 1, \quad \forall k\in\mathcal K,\\
& \boldsymbol Q_k+{\sum_{i=1,i\neq k}^K \alpha_i \boldsymbol g_{ik}\boldsymbol g_{ik}^H} \succeq \frac{\alpha_k}{\Gamma_k}\boldsymbol g_{kk}\boldsymbol g_{kk}^H,   \quad  \forall k\in\mathcal K,\\
& \alpha_k\geq 0, \quad \forall k\in\mathcal K,
\end{align}
\end{subequations}
where  $\boldsymbol q\!=\![q_{11},\cdots,q_{1N},\cdots,q_{KN}]^T$, $\boldsymbol \alpha\!=\![\alpha_{1},\cdots,\alpha_{K}]^T$,
$q_{kn}$ and $\alpha_k$ are the Lagrange multipliers corresponding to power constraints (\ref{Alo3max1}b) and SINR constraints (\ref{Alo3max1}a), respectively,
and $\boldsymbol Q_k$ is defined in \eqref{Alo3max1dueq1_1}.
\end{theorem}

\itshape {Proof:}  \upshape
The Lagrangian function for the optimization
problem (\ref{Alo3max1}) is given by:
\begin{align}\label{Alo3max1dueq1}
\mathcal L (\boldsymbol w, \boldsymbol p,\boldsymbol q,  \boldsymbol \alpha)= &\sum_{k=1}^K  p_k
+\sum_{k=1}^K \sum_{n=1}^N q_{kn}([\boldsymbol w_k\boldsymbol w_k^H]_{n,n}-p_k)+
\nonumber\\&
-\sum_{k=1}^K \alpha_k\left(\frac{1}{\Gamma_k} \left| \boldsymbol g_{kk}^H\boldsymbol w_k \right|^2-
{\sum_{i=1,i\neq k}^K   \left|\boldsymbol g_{ki}^H\boldsymbol w_i  \right|^2-\sigma^2}
\right).
\end{align}

Denoting
\begin{equation}\label{Alo3max1dueq1_1}
\boldsymbol Q_k=\text{diag}(q_{k1},\cdots,q_{kN}),
\end{equation}
 we can rewrite Lagrangian function \eqref{Alo3max1dueq1} by:
\begin{align}\label{Alo3max1dueq2}
\mathcal L(\boldsymbol w, \boldsymbol p,\boldsymbol q,  \boldsymbol \alpha)= &
\sum_{k=1}^K  \alpha_k \sigma^2-\sum_{k=1}^K  p_k\left(\sum_{n=1}^N q_{kn}-1\right)
\nonumber\\&
+\sum_{k=1}^K \boldsymbol w_k^H\left( \boldsymbol Q_k+{\sum_{i=1,i\neq k}^K \alpha_i \boldsymbol g_{ik}\boldsymbol g_{ik}^H}
-\frac{\alpha_k}{\Gamma_k}\boldsymbol g_{kk}\boldsymbol g_{kk}^H
\right)\boldsymbol w_k.
\end{align}
The dual objective can be given by \cite{boyd2004convex}:
\begin{align}\label{Alo3max1dueq2_1}
D(\boldsymbol q,  \boldsymbol \alpha)=\min_{\boldsymbol w, \boldsymbol p}\mathcal L &(\boldsymbol w, \boldsymbol p,\boldsymbol q,  \boldsymbol \alpha).
\end{align}

Since $p_k$ must be positive and there are no constraints on the
beamforming $\boldsymbol w_k$, we have $D(\boldsymbol q,  \boldsymbol \alpha)=-\infty$ if $\sum_{n=1}^Nq_{kn}\geq 1$ or $\boldsymbol Q_k+{\sum_{i=1,i\neq k}^K \alpha_i \boldsymbol g_{ik}\boldsymbol g_{ik}^H} -\frac{\alpha_k}{\Gamma_k}\boldsymbol g_{kk}\boldsymbol g_{kk}^H$ is not positive semidefinite.

Due to the fact that $\boldsymbol q$ and  $\boldsymbol \alpha$ should be selected that the dual objective is finite.
As a result, constraints that $\sum_{n=1}^Nq_{kn}\leq 1$ and $\boldsymbol Q_k+{\sum_{i=1,i\neq k}^K \alpha_i \boldsymbol g_{ik}\boldsymbol g_{ik}^H} -\frac{\alpha_k}{\Gamma_k}\boldsymbol g_{kk}\boldsymbol g_{kk}^H$ is positive semidefinite should be satisfied.
Formally, the Lagrangian dual problem can be stated as \eqref{Alo3max2}.
\hfill $\Box$

Since the objective is linear and the constraints are either linear or linear matrix inequalities,
dual problem \eqref{Alo3max2} is a semidefinite programming (SDP) problem, which can be solved by using the standard CVX toolbox \cite{grant2008cvx,yang2019eeFL}.
Having obtained the dual variables by solving dual problem \eqref{Alo3max2}, it remains to obtain the optimal beamforming $\boldsymbol w$ and transmit power $\boldsymbol p$.
To find the optimal $\boldsymbol w$, we calculate the gradient of the
Lagrangian function for problem (\ref{Alo3max1}) with respect to $\boldsymbol w$
and set it to zero:
\begin{align}\label{Alo3max1dueq2_2}
\frac{\partial \mathcal L(\boldsymbol w, \boldsymbol p,\boldsymbol q,  \boldsymbol \alpha)}{\partial\boldsymbol w_k}
=&2\Bigg( \boldsymbol Q_k+{\sum_{i=1,i\neq k}^K \alpha_i \boldsymbol g_{ik}\boldsymbol g_{ik}^H}
-\frac{\alpha_k}{\Gamma_k}\boldsymbol g_{kk}\boldsymbol g_{kk}^H
\Bigg)\boldsymbol w_k=\boldsymbol 0.
\end{align}
Based on \eqref{Alo3max1dueq2_2}, we have:
\begin{align}\label{Alo3max1dueq2_3}
\left( \boldsymbol Q_k+{\sum_{i=1}^K \alpha_i \boldsymbol g_{ik}\boldsymbol g_{ik}^H}
\right)\boldsymbol w_k=\frac{1+\Gamma_k}{\Gamma_k}\alpha_k\boldsymbol g_{kk}\boldsymbol g_{kk}^H \boldsymbol w_k.
\end{align}
Solving equation \eqref{Alo3max1dueq2_3} yields:
\begin{align}\label{Alo3max1dueq2_5}
\boldsymbol w_k=\left( \boldsymbol Q_k+{\sum_{i=1}^K \alpha_i \boldsymbol g_{ik}\boldsymbol g_{ik}^H}
\right)^{\dagger}\frac{1+\Gamma_k}{\Gamma_k}\alpha_k\boldsymbol g_{kk}\boldsymbol g_{kk}^H \boldsymbol w_k.
\end{align}
Since $\boldsymbol g_{kk}^H \boldsymbol w_k$ is a scalar, the optimal $\boldsymbol w$ has the following expression:
\begin{align}\label{Alo3max1dueq3_1}
\boldsymbol w_k^\star={\sqrt{p_k^\star}} \boldsymbol\theta_k^\star,
\end{align}
where
\begin{align}\label{Alo3max1dueq3_2_2}
\boldsymbol \theta_k^\star =\frac{ \sqrt{N}{\left( \boldsymbol Q_k+{\sum_{i=1}^K \alpha_i \boldsymbol g_{ik}\boldsymbol g_{ik}^H}
\right)^{\dagger}\boldsymbol g_{kk}} }{\left\| {\left( \boldsymbol Q_k+{\sum_{i=1}^K \alpha_i \boldsymbol g_{ik}\boldsymbol g_{ik}^H}
\right)^{\dagger}\boldsymbol g_{kk}} \right\|}.
\end{align}
Note that the $\ell_2$-norm of phase vector $\boldsymbol \theta_k^\star$ is $\|\boldsymbol \theta_k^\star\|=\sqrt{N}$ since the  module of each element in phase vector is unit.
To obtain the value of power $p_k^\star$, we find that the minimum SINR constraints (\ref{Alo3max1}a) must hold with equality for all users at the optimum solution.
Substituting \eqref{Alo3max1dueq3_1} into SINR constraints (\ref{Alo3max1}a) and setting them with equality, we can obtain:
\begin{equation}\label{Alo3max1dueq5}
 \frac{ \left| \boldsymbol g_{kk}^H\boldsymbol \theta_k^\star \right|^2}{ \Gamma_k} p_k^\star=
{\sum_{i=1,i\neq k}^K   \left|\boldsymbol g_{ki}^H\boldsymbol \theta_i^\star  \right|^2p_i^\star+\sigma^2}.
\end{equation}
By using the concept of standard interference function, equation \eqref{Alo3max1dueq5} can be written in the following form:
\begin{equation}\label{Alo3max1dueq5_1}
 \boldsymbol  p^\star= \boldsymbol f(\boldsymbol  p^\star),
\end{equation}
where $\boldsymbol f=[f_1,\cdots,f_K]^T$ and
\begin{equation}\label{Alo3max1dueq5_2}
 f_k(\boldsymbol  p^\star)= \frac{ \Gamma_k}{ \left| \boldsymbol g_{kk}^H\boldsymbol \theta_k^\star \right|^2}
\left({\sum_{i=1,i\neq k}^K   \left|\boldsymbol g_{ki}^H\boldsymbol \theta_i^\star  \right|^2p_i^\star+\sigma^2}\right).
\end{equation}
By checking the positivity, monotonicity, and scalability properties, we can prove that function $\boldsymbol f(\boldsymbol  p^\star)$ is always a standard interference function \cite{yates1995framework}, which allows us to use the iterative power control scheme to solve equation  \eqref{Alo3max1dueq5_1}.
The iterative power control scheme is given by:
\begin{equation}\label{Alo3max1dueq5_3}
 \boldsymbol  p^{(t)}= \boldsymbol f(\boldsymbol  p^{(t-1)}),
\end{equation}
where the superscript $(t)$ means the value of the variable in the $t$-th iteration.
According to \cite[Theorem 2]{yates1995framework}, the iterative power control scheme \eqref{Alo3max1dueq5_3} always converges to the unique fixed point $\boldsymbol  p^\star$ if \eqref{Alo3max1dueq5_1} is feasible.

The dual method for solving problem \eqref{Alo3max2} is summarized in Algorithm 1.
Since the duality gap is zero for the large number of RIS units, the  solution $(\boldsymbol w_k^\star, p_k^{\star})$ obtained by the dual method in Algorithm 1  is the optimal solution of the original problem in \eqref{Alo3max1}.

\begin{algorithm}[t] \vspace{-0mm}
\caption{Dual Method for Problem \eqref{Alo3max1}}
\begin{algorithmic}[1]\label{singleuseropiAlo}
\STATE Solve the dual problem \eqref{Alo3max2} by using SDP.
\STATE Calculate the optimal phase beamforming vector $\boldsymbol \theta_k^\star$ according to \eqref{Alo3max1dueq3_2_2}.
\STATE Initialize $\boldsymbol p^{(0)}=\boldsymbol 0$, iteration number $t=1$, and set the accuracy $\epsilon$.
\REPEAT
\FOR{$k=1:K$}
\STATE Update $ p_k^{(t)}= \frac{ \Gamma_k}{ \left| \boldsymbol g_{kk}^H\boldsymbol \theta_k^\star \right|^2}
\left({\sum_{i=1,i\neq k}^K   \left|\boldsymbol g_{ki}^H\boldsymbol \theta_i^\star  \right|^2p_i^{(t-1)}+\sigma^2}\right)$.
\ENDFOR
\STATE Set $t=t+1$ and $\boldsymbol p^{(t)}=[p_1^{(t)},\cdots, p_K^{(t)}]^T$.
\UNTIL  $\|\boldsymbol  p^{(t)}-\boldsymbol f(\boldsymbol  p^{(t)})\|<\epsilon$.
\STATE Output $\boldsymbol w_k^\star={\sqrt{p_k^{(t)}}} \boldsymbol\theta_k^\star$, $p_k^\star=\sqrt{p_k^{(t)}}$, $\forall k \in \mathcal K$.
\end{algorithmic} \vspace{-0mm}
\end{algorithm}

{\color{myc1}
\subsection{SDR Approach}
In this section, we apply the semidefinite
relaxation (SDR) technique to solve problem \eqref{Alo3max1}.
We introduce matrix $\boldsymbol W_k=\boldsymbol w_k\boldsymbol w_k^H$, which needs to satisfy $\boldsymbol W_k\succeq0$ and the rank of $\boldsymbol W_k$ is one.
As a result, problem \eqref{Alo3max1} is equivalent to
\begin{subequations}\label{re2Alo3max1eq1}
\begin{align}
\mathop{\min}_{ \boldsymbol W,    \boldsymbol p} \:&
 \sum_{k=1}^K  p_k \tag{\theequation}\\
\textrm{s.t.} \:
&  {  \boldsymbol g_{kk}^H\boldsymbol W_k \boldsymbol g_{kk}}
\geq \Gamma_k\left({\sum_{i=1,i\neq k}^K   \boldsymbol g_{ki}^H\boldsymbol W_i  \boldsymbol g_{ki}+\sigma^2}\right), \quad \forall k\in\mathcal K,\\
&  [\boldsymbol W_k]_{nn}=p_k,   \quad  \forall k\in\mathcal K, n\in\mathcal N, \\
&  rank(\boldsymbol W_k)=1, \quad  \forall k\in\mathcal K,    \\
& \boldsymbol W_k \succeq0, \quad  \forall k\in\mathcal K,
\end{align}
\end{subequations}
where $\boldsymbol W=[\boldsymbol W_1,\boldsymbol W_2,\cdots,\boldsymbol W_K]$.
However. \eqref{re2Alo3max1eq1} is still a
nonconvex problem due to the nonconvex rank-one constraint (\ref{re2Alo3max1eq1}c). We apply SDR to relax this rank-one constraint
  and problem \eqref{re2Alo3max1eq1} can be represented as:
\begin{subequations}\label{re2Alo3max1eq2}
\begin{align}
\mathop{\min}_{ \boldsymbol W,    \boldsymbol p} \:&
 \sum_{k=1}^K  p_k \tag{\theequation}\\
\textrm{s.t.} \:
&  {  \boldsymbol g_{kk}^H\boldsymbol W_k \boldsymbol g_{kk}}
\geq \Gamma_k\left({\sum_{i=1,i\neq k}^K   \boldsymbol g_{ki}^H\boldsymbol W_i  \boldsymbol g_{ki}+\sigma^2}\right), \quad \forall k\in\mathcal K,\\
&  [\boldsymbol W_k]_{nn}=p_k,   \quad  \forall k\in\mathcal K, n\in\mathcal N, \\
& \boldsymbol W_k \succeq0, \quad  \forall k\in\mathcal K.
\end{align}
\end{subequations}
Problem \eqref{re2Alo3max1eq2} is a standard convex problem, which can be
solved by existing toolboxes such as CVX. Since the rank-one constraint
is relaxed, the solution of problem \eqref{re2Alo3max1eq2} is not necessarily to
the optimal solution of the original problem \eqref{re2Alo3max1eq2},
which implies that the optimal objective value
of problem \eqref{re2Alo3max1eq2} only serves an upper bound of problem \eqref{re2Alo3max1eq1}. Thus, we can use the method in  \cite{wu2018intelligent}
to construct a rank-one solution from the obtained higher-rank solution to problem \eqref{re2Alo3max1eq1}.

According to \cite{so2007approximating}, the
SDR approach followed by a sufficiently large number of randomizations guarantees at least a
$\frac{\pi}{4}$-approximation of the optimal objective value of problem \eqref{re2Alo3max1eq1}.
Since constructing a feasible rank-one solution requires a lot of randomizations, the complexity of the SDR approach is higher than the proposed dual method. Besides, the proposed dual method can be guaranteed to be the optimal for large number of users or RIS unit elements.}

\subsection{MRT and ZF Beamforming}
To solve the sum power minimization problem \eqref{sys2max1} in the conventional ways, we provide the MRT and ZF beamforming methods.
Note that the optimal power control can be obtained by using the iterative power control scheme \eqref{Alo3max1dueq5_3}, this section only provides the method of optimizing the phase beamforming $\boldsymbol \theta_k$.
\subsubsection{MRT beamforming}
In MRT beamforming, the beamforming is designed such that the received signal at each user is maximized.
Mathematically, the MRT beamforming problem is formulated as:
\begin{subequations}\label{mrtmax1}
\begin{align}
\mathop{\max}_{ \boldsymbol\theta_k} \quad&
|\boldsymbol g_{kk}^H \boldsymbol \theta_k| \tag{\theequation}\\
\textrm{s.t.} \quad
&|\theta_{kn}| =1,   \quad  \forall  n\in\mathcal N.
\end{align}
\end{subequations}
To solve problem \eqref{mrtmax1}, the MRT beamforming can be expressed as:
\begin{equation}\label{MRT}
\theta_{kn}^\star=\frac{[\boldsymbol g_{kk}]_n}{|[\boldsymbol g_{kk}]_n|}, \quad \forall n \in\mathcal N.
\end{equation}

From \eqref{MRT}, we can see that the optimal phase vector $\boldsymbol\theta_k^\star$ should be tuned such that the signal that passes through all RIS units is aligned to be a signal vector with the equal phase at each element, i.e., $\theta_{kn}^\star [\boldsymbol g_{kk}]_n$ is a real number for all $n$.

\subsubsection{ZF bemaforming}
The idea of ZF beamforming is to invert the channel matrix at the transmitter in order to create orthogonal channels
between the BS and users.
The ZF beamforming problem can be formulated as:
\begin{subequations}\label{zfmax1}
\begin{align}
\mathop{\max}_{ \boldsymbol\theta_k} \quad&
 |\boldsymbol g_{kk}^H \boldsymbol \theta_k| \tag{\theequation}\\
\textrm{s.t.} \quad
&\boldsymbol g_{ik}^H \boldsymbol \theta_k=0, \quad \forall i\neq k,\\
&|\theta_{kn}| =1,   \quad  \forall  n\in\mathcal N.
\end{align}
\end{subequations}

Due to the nonconvex unit-modulus constraints (\ref{zfmax1}b), the conventional ZF method cannot be directly applied.
With new nonconvex constraints (\ref{zfmax1}b), it is of importance to investigate the feasibility of (\ref{zfmax1}).
For the feasibility of problem \eqref{zfmax1}, we have the following lemma.
\begin{lemma}
Problem \eqref{zfmax1} is feasible only if the following constraints are satisfied:
\begin{equation}\label{MRTZFeq00}
2\max_{n\in\mathcal N}|[\boldsymbol g_{ki}]_n|\leq\sum_{n\in\mathcal N} |[\boldsymbol g_{ki}]_n|, \quad \forall i \neq k.
\end{equation}
\end{lemma}
\itshape {Proof:}  \upshape
According to constraints (\ref{zfmax1}a), we have:
\begin{equation}\label{MRTZFeq0}
[\boldsymbol g_{ki}]_l  \theta_{kl}^*=-\sum_{n\in\mathcal N,n\neq l} [\boldsymbol g_{ki}]_n \theta_{kn}^* , \quad \forall i \neq k.
\end{equation}
Taking the absolute value at both sides yields:
\begin{equation}\label{MRTZFeq0_1}
|[\boldsymbol g_{ki}]_l \theta_{kl}^*|=\left |\sum_{n\in\mathcal N,n\neq l} [\boldsymbol g_{ki}]_n \theta_{kn}^*\right|.
\end{equation}
Further combining constraints (\ref{zfmax1}b), we have:
\begin{align}\label{MRTZFeq0_2}
|[\boldsymbol g_{ki}]_l |&=\left|\sum_{n\in\mathcal N,n\neq l} [\boldsymbol g_{ki}]_n \theta_{kn}^*\right|
\leq \sum_{n\in\mathcal N,n\neq l} |[\boldsymbol g_{ki}]_n \theta_{kn}^*|=\sum_{n\in\mathcal N,n\neq l} |[\boldsymbol g_{ki}]_n|,
\end{align}
where the inequality follows from the triangle inequality.
Adding $|[\boldsymbol g_{ki}]_l |$ on both sides, \eqref{MRTZFeq0_2} becomes:
\begin{align}\label{MRTZFeq0_3}
2|[\boldsymbol g_{ki}]_l |&\leq \sum_{n\in\mathcal N} |[\boldsymbol g_{ki}]_n|.
\end{align}
Since inequality \eqref{MRTZFeq0_3} should be satisfied for any $l\in\mathcal N$, conditions \eqref{MRTZFeq00} are obtained by using the $\max_{l\in\mathcal N}$ operation on both sides of \eqref{MRTZFeq0_3}.
\hfill $\Box$

\begin{figure}[t]
\centering
\includegraphics[width=3.0in]{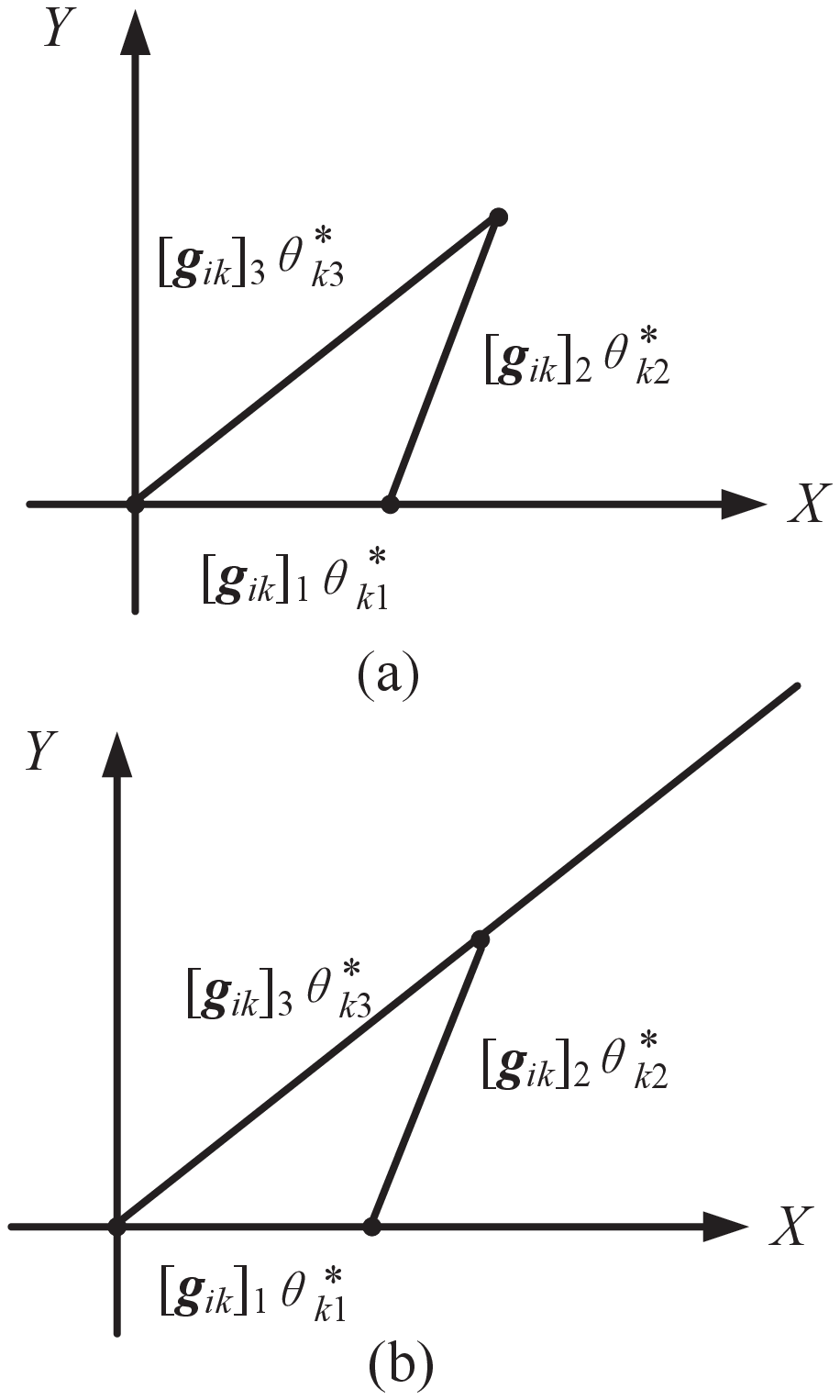}
\vspace{-2em}
\caption{An example of $N=3$ and $\max_{n\in\mathcal N}|[\boldsymbol g_{ki}]_n|=|[\boldsymbol g_{ki}]_3|$. (a) $|[\boldsymbol g_{ki}]_3|\leq |[\boldsymbol g_{ki}]_1| + |[\boldsymbol g_{ki}]_2|$ and (b) $|[\boldsymbol g_{ki}]_3|>|[\boldsymbol g_{ki}]_1| + |[\boldsymbol g_{ki}]_2|$.} \label{figsys3}
\vspace{-2em}
\end{figure}

{\color{myc1}To elaborate lemma 2 further,
an example of $N=3$ and $\max_{n\in\mathcal N}|[\boldsymbol g_{ki}]_n|=|[\boldsymbol g_{ki}]_3|$  is shown in Fig. \ref{figsys3}.
In this figure, we consider the special case that there are $N=3$ reflecting elements and the channel gain with the maximum amplitude is $\max_{n\in\mathcal N}|[\boldsymbol g_{ki}]_n|=|[\boldsymbol g_{ki}]_3|$.
If the condition \eqref{MRTZFeq00} is satisfied, we have
\begin{equation*}
2\max_{n\in\mathcal N}|[\boldsymbol g_{ki}]_n|=2|[\boldsymbol g_{ki}]_3|\leq\sum_{n=1}^3 |[\boldsymbol g_{ki}]_n|,
\end{equation*}
i.e.,
\begin{equation*}
 |[\boldsymbol g_{ki}]_3|\leq  |[\boldsymbol g_{ki}]_1|+ |[\boldsymbol g_{ki}]_2|,
\end{equation*}
as shown in Fig. \ref{figsys3}(a).
In this case, we can always construct a triangle with the length of three edges are respectively $|[\boldsymbol g_{ki}]_1|$, $|[\boldsymbol g_{ki}]_2|$, and $|[\boldsymbol g_{ki}]_3|$, which means that there always exists $\boldsymbol \theta_k$ such that $\sum_{n=1}^N \theta_{kn}^*[\boldsymbol g_{ki}]_n=0$.
However, if the condition \eqref{MRTZFeq00} is not satisfied, i.e., $ |[\boldsymbol g_{ki}]_3| > |[\boldsymbol g_{ki}]_1|+ |[\boldsymbol g_{ki}]_2|$, as shown in Fig. \ref{figsys3}(b).
In this case, we cannot construct a triangle with the length of three edges that are respectively $|[\boldsymbol g_{ki}]_1|$, $|[\boldsymbol g_{ki}]_2|$, and $|[\boldsymbol g_{ki}]_3|$. As a result, there is no solution for equation $\sum_{n=1}^N \theta_{kn}^*[\boldsymbol g_{ki}]_n=0$.}

After checking the feasibility, we find that ZF beamforming problem (\ref{zfmax1}) is still difficult to solve (i.e., it is nonconvex) because of nonconvex constraints (\ref{zfmax1}b).
In the following, an iterative algorithm is proposed to effectively solve ZF beamforming problem (\ref{zfmax1}) with low complexity.

Without loss of generality, the term $\boldsymbol g_{kk}^H \boldsymbol \theta_k$ in the objective function (\ref{zfmax1}) can be expressed as a real number through an arbitrary rotation to phase beamforming $\boldsymbol \theta_k$.
As a result, the objective function (\ref{zfmax1}) can be equivalent to
\begin{equation}\label{MRTZFeq1_0}
\mathop{\max}_{ \boldsymbol\theta_k} \quad
\mathcal R( \boldsymbol g_{kk}^H \boldsymbol \theta_k).
\end{equation}

According to constraint (\ref{zfmax1}a),
$\boldsymbol \theta_k$ must lie in the orthogonal complement of the subspace $\text{span}\{\boldsymbol g_{ik}, \forall i \neq k\}$  and the orthogonal projector matrix on this  orthogonal complement is \cite{6425521}
\begin{equation}\label{MRTZFeq1}
\boldsymbol Z_k=\boldsymbol I -\boldsymbol G_k(\boldsymbol G_k^H\boldsymbol G_k)^{\dagger}\boldsymbol G_k^H,
\end{equation}
where
\begin{equation}\label{MRTZFeq1_2}
\boldsymbol G_k=[\boldsymbol g_{1k},\cdots,\boldsymbol g_{(k-1)k},\boldsymbol g_{(k+1)k}\cdots,\boldsymbol g_{Kk}].
\end{equation}
For any vector $\boldsymbol \theta_k$ satisfying constraint (\ref{zfmax1}a), $\boldsymbol \theta_k$ can be expressed by:
\begin{equation}\label{MRTZFeq1_3}
\boldsymbol \theta_k= \boldsymbol Z_k \boldsymbol v_k,
\end{equation}
where $\boldsymbol v_k$ is an  $N\times1$ complex  vector to be optimized.

Based on \eqref{MRTZFeq1_0} and \eqref{MRTZFeq1_3}, ZF beamforming optimization problem \eqref{zfmax1} is equivalent to:
\begin{subequations}\label{zfmax2}
\begin{align}
\mathop{\max}_{ \boldsymbol\theta_k, \boldsymbol v_k} \quad&
\mathcal R( \boldsymbol g_{kk}^H \boldsymbol \theta_k) \tag{\theequation}\\
\textrm{s.t.} \quad
&\boldsymbol \theta_k= \boldsymbol Z_k \boldsymbol v_k,\\
&|\theta_{kn}| =1,   \quad  \forall  n\in\mathcal N.
\end{align}
\end{subequations}



Since both variables $\boldsymbol \theta_k$ and $\boldsymbol v_k$ are coupled in the constraint (\ref{zfmax2}a), we introduce the barrier method to transform problem \eqref{zfmax2} as follows
\begin{subequations}\label{zfmax3_1}
\begin{align}
\mathop{\max}_{ \boldsymbol\theta_k, \boldsymbol v_k} \quad&
\mathcal R( \boldsymbol g_{kk}^H \boldsymbol Z_k \boldsymbol v_k) -\lambda\|\boldsymbol \theta_k-\boldsymbol Z_k \boldsymbol v_k\|^2 \tag{\theequation}\\
\textrm{s.t.} \quad
&|\theta_{kn}| =1,   \quad  \forall  n\in\mathcal N,
\end{align}
\end{subequations}
where $\lambda>0$ is a large penalty factor \cite{boyd2004convex}.
To solve problem \eqref{zfmax3_1},
an iterative algorithm is proposed via
alternatingly optimizing $\boldsymbol \theta_k$ with fixed $\boldsymbol v_k$, and updating $\boldsymbol v_k$ with optimized $\boldsymbol \theta_k$ in the previous step, which admits efficient closed-form solutions in each step.

In the first step, given $\boldsymbol v_k$, problem \eqref{zfmax3_1} can be formulated as:
\begin{subequations}\label{zfmax3_2}
\begin{align}
\mathop{\max}_{ \boldsymbol\theta_k} \quad&
 -\|\boldsymbol \theta_k-\boldsymbol Z_k \boldsymbol v_k\|^2
 =2\boldsymbol \theta_k \boldsymbol Z_k \boldsymbol v_k-N-\boldsymbol v_k^H\boldsymbol Z_k^H\boldsymbol Z_k \boldsymbol v_k\tag{\theequation}\\
\textrm{s.t.} \quad
&|\theta_{kn}| =1,   \quad  \forall  n\in\mathcal N.
\end{align}
\end{subequations}
To maximize \eqref{zfmax3_2}, it is easy to get
\begin{equation}\label{zfmax3_2eq1}
 \theta_{kn}^\star=\frac{[\boldsymbol Z_k \boldsymbol v_k]_n^*}{|[\boldsymbol Z_k \boldsymbol v_k]_n|}, \quad \forall n \in\mathcal N.
\end{equation}

In the second step, we update the value of $\boldsymbol v_k$ with the optimized $\boldsymbol \theta_k$ in  \eqref{zfmax3_2eq1}.
Then, problem \eqref{zfmax3_1} becomes
\begin{subequations}\label{zfmax3_5}
\begin{align}
\mathop{\max}_{\boldsymbol v_k} \quad&
\mathcal R( \boldsymbol g_{kk}^H \boldsymbol Z_k \boldsymbol v_k) -\lambda\|\boldsymbol \theta_k-\boldsymbol Z_k \boldsymbol v_k\|^2. \tag{\theequation}
\end{align}
\end{subequations}
Problem \eqref{zfmax3_5} is convex and thus the optimal solution can be obtained by setting the gradient to zero.
We calculate the gradient of
 (\ref{zfmax3_5}) with respect to $\boldsymbol v_k$
and set it to zero, i.e.,
\begin{equation}\label{zfmax3_5eq1}
\mathcal R(\boldsymbol Z_k^H \boldsymbol g_{kk}+2\lambda\boldsymbol Z_k^H(\boldsymbol \theta_k-\boldsymbol Z_k \boldsymbol v_k))=\boldsymbol 0,
\end{equation}
which gives
\begin{equation}\label{zfmax3_5eq2}
\boldsymbol v_k=(\boldsymbol Z_k^H\boldsymbol Z_k)^\dagger
\left(\frac{1}{2\lambda}\boldsymbol Z_k^H \boldsymbol g_{kk}+\boldsymbol Z_k^H \boldsymbol \theta_k\right).
\end{equation}

Note that the iterative algorithm for solving problem \eqref{zfmax3_1} is summarized in Algorithm \ref{zfAlo}.
Due to the fact that the optimal solution of problem \eqref{zfmax3_2} and \eqref{zfmax3_5} can be obtained, the objective value (\ref{zfmax3_1}) is always increasing at each iteration.
Since the objective value of  problem \eqref{zfmax3_1} is increasing at each iteration and  the objective value of problem \eqref{zfmax3_1} always has a finite upper bound,
 Algorithm~\ref{zfAlo} always converges.

\begin{algorithm}[t]
\caption{Iterative Optimization for Problem \eqref{zfmax3_1}}
\begin{algorithmic}[1]\label{zfAlo}
\STATE Initialize $\boldsymbol v_k^{(0)}$. Set iteration number $t=1$.
\REPEAT
\STATE Given $\boldsymbol v_k^{(t-1)}$, obtain the optimal solution of problem \eqref{zfmax3_2}, which is denoted by $\boldsymbol \theta_k^{(t)}$.
\STATE Given $\boldsymbol \theta_k^{(t)}$, obtain the optimal solution of problem \eqref{zfmax3_5}, which is denoted by $\boldsymbol v_k^{(t)}$.
\STATE Set $t=t+1$.
\UNTIL the objective value (\ref{zfmax3_1}) converges.
\end{algorithmic}
\end{algorithm}


\subsection{Power Scaling Law with Infinite RIS Units}
We characterize the scaling law of the average received power at the user with respect to
the number of RIS units, i.e., $N \rightarrow \infty$.
For simplicity, we consider the single-user case with $K=1$.
The received power at the user is $P=P_0|\boldsymbol g^T \boldsymbol \theta|^2$, where $P_0$ is the transmitted power at the BS, 
$\boldsymbol g$ is the channel gain between the RIS and the user, and $\boldsymbol \theta$ is the phase beamfoming of the BS.
We compare two phase beamforming solutions: (i) $\boldsymbol \theta =\boldsymbol 1$ and (ii) the MRT beamforming $[\boldsymbol \theta]_n =\frac{[\boldsymbol g]_n}{|[\boldsymbol g]_n|}$.

\begin{theorem}
Assume that $\boldsymbol g\sim\mathcal {CN}(\boldsymbol 0, \rho\boldsymbol I)$.
If $N\rightarrow \infty$, we have:
\begin{equation}\label{sc1umax5_0eq1}
P=\left\{ \begin{array}{ll}
\!\!N\rho P_0, &\text{if}\; \boldsymbol \theta =\boldsymbol 1, \\
\!\!\frac{(\pi^2-7\pi+16)}{4}N^2\rho P_0, &\text{if}\;  [\boldsymbol \theta]_n =\frac{[\boldsymbol g]_n}{|[\boldsymbol g]_n|}, \forall n \in\mathcal N.
\end{array} \right.
\end{equation}
\end{theorem}

\itshape {Proof:}  \upshape
When $\boldsymbol \theta =\boldsymbol 1$, we have $\boldsymbol g^T \boldsymbol \theta\sim\mathcal {CN}(0,N\rho)$.
Thus, we have $P=\mathbb E(P_0|\boldsymbol g^T \boldsymbol \theta|^2)=N\rho P_0$.

When $[\boldsymbol \theta]_n =\frac{[\boldsymbol g]_n}{|[\boldsymbol g]_n|}$, we have
\begin{equation}
\boldsymbol g^T \boldsymbol \theta=\sum_{n=1}^N [\boldsymbol \theta]_n^* [\boldsymbol g]_n = \sum_{n=1}^N |[\boldsymbol g]_n|.
\end{equation}
Since $\boldsymbol g\sim\mathcal {CN}(\boldsymbol 0, \rho\boldsymbol I)$, $|[\boldsymbol g]_n|$ follows the Rayleigh distribution with mean $\frac{\sqrt {\pi\rho} }{2}$ and variance $\frac{(4-\pi)\rho}{2}$.
According to the central limit theorem, $\sum_{n=1}^N |[\boldsymbol g]_n|\sim\mathcal {CN}(\frac{N\sqrt {\pi\rho} }{2},\frac{N(4-\pi)\rho}{2})$.
As a result, we have
\begin{equation}
P=\mathbb E(P_0|\boldsymbol g^T \boldsymbol \theta|^2)=\frac{(\pi^2-7\pi+16)}{4}N^2\rho P_0.
\end{equation}
The proof completes.
\hfill $\Box$

According to Theorem 2, it is obtained that the received power only linearly increase with the number of RIS units if there is no optimization on the phase beamforming.
However, under the proposed phase beamforming, the received power quadratically increase with the number of RIS units, which can greatly increase the received signal strength especially for large number of RIS units.
For the conventional massive
MIMO, the order of transmit beamforming gain is $N$ \cite{ngo2013energy,wu2019intelligent}.
{\color{myc2}The gain of RIS as a transmitter lies in the fact that RIS is used to modulate the signal via reflection  and increasing the number of RIS elements, which does not require any additional transmit power of the transmitter.}
\subsection{Complexity Analysis}
The complexity of Algorithm 1 lies in solving the SDP problem \eqref{Alo3max2}.
Since the total number of variables is $S_1=(K+1)N$ and
 the total number of constraints is $S_2=3K$, the complexity of solving  problem \eqref{Alo3max2} is $\mathcal O (S_1^2 S_2)=\mathcal O (K^2N^3)$ according to \cite{lobo1998applications}.

The complexity of Algorithm 2 lies in solving problem \eqref{zfmax3_5} at each step.
According to \eqref{zfmax3_5eq2}, the complexity of solving problem \eqref{zfmax3_5} is $\mathcal O(N^3)$.
The total complexity of Algorithm 2 is $\mathcal O(IN^3)$, where $I$ denotes the number of iterations for Algorithm 2.

\section{Numerical Results}
In this section, we evaluate the performance of the proposed algorithm.
There are $K$ users  uniformly distributed in a square area of size $500$ m $\times$ $500$~m  with the BS located at its center.
The large-scale pathloss model is $ 10^{-3.76} d^{-\alpha}$ ($d$ is in m), where $\alpha$ is the pathloss factor.
The noise power is -114 dBm.
For the channel gain $\boldsymbol g_{ki}$, we set $[\boldsymbol g_{ki}]_n\sim\mathcal {CN}(0,1)$, $\forall k, i\in\mathcal K, n \in \mathcal N$ \cite{8846706,chen2019joint}.
Unless specified otherwise, we choose a pathloss factor $\alpha=3$, {\color{myc1}a total of $K=8$ users}, a number of $N=20$ RIS units allocated for each user, a penalty factor $\lambda=10^3$, and an equal SINR requirement $\Gamma_1=\cdots\Gamma_K=\Gamma=2$.
Additionally, the effectiveness of the proposed dual method (labeled as ``DM'') is verified by comparing with the MRT and ZF methods.

\begin{figure}[t]
\centering
\includegraphics[width=3.5in]{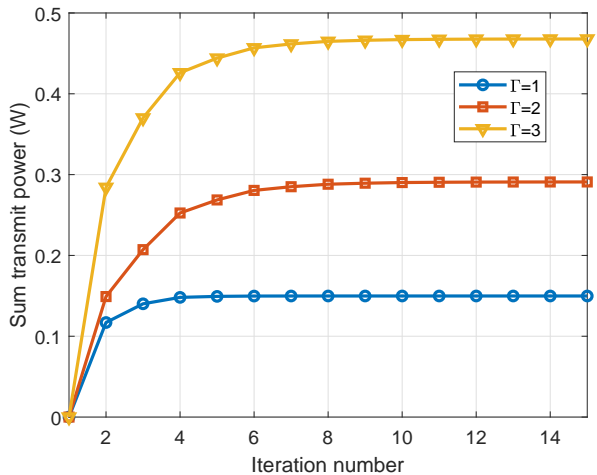}
\vspace{-1em}
\caption{Convergence behaviour of Algorithm \ref{singleuseropiAlo} under different SINR requirements.} \label{fig1}
\vspace{-1.5em}
\end{figure}

Fig. \ref{fig1} illustrates the convergence of Algorithm \ref{singleuseropiAlo} under different SINR requirements.
It can be seen that the proposed algorithm converges fast, and six iterations
are sufficient to converge, which shows the effectiveness of the proposed algorithm in terms of convergence performance.

\begin{figure}[t]
\centering
\includegraphics[width=3.5in]{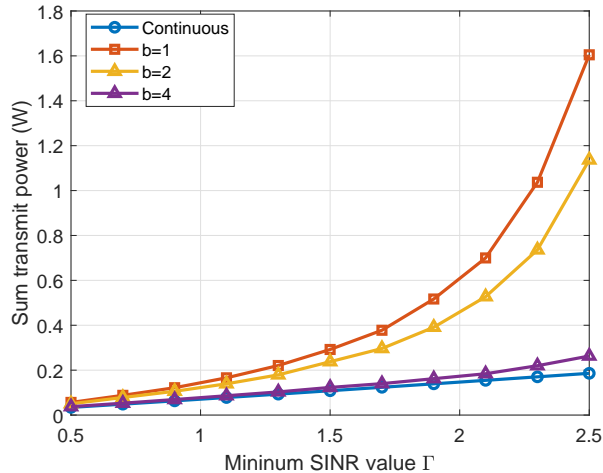}
\vspace{-1em}
\caption{Sum transmit power versus the minimum SINR requirement for continuous and discrete phase shift schemes.} \label{fig311}
\vspace{-1.5em}
\end{figure}
{\color{myc1}The sum transmit power versus the minimum SINR requirement for continuous and discrete phase shift schemes is shown in Fig. \ref{fig311}.
In this figure, $b$ denotes the
number of bits used to indicate the number of phase shift levels $L$ where $L = 2^b$. For simplicity,
we assume that such discrete phase-shift values are obtained by uniformly quantizing the interval $[0, 2 \pi)$. Thus, the set of discrete phase-shift values at each element is given by
\begin{equation}
\mathcal F=\left\{0, \frac{2\pi}{L},\frac{4\pi}{L}, \cdots, \frac{2(L-1)\pi}{L}\right\}.
\end{equation}
Denote $\boldsymbol \theta^*$ as the obtained result of considering continuous phase shifts.
We use the rounding method to obtain discrete phase shifts solution $\hat{\boldsymbol \theta}$, where
\begin{equation}
\hat \theta_{kn}=\arg \min_{\theta_{kn}\in\mathcal F} |\theta_{kn}-\theta_{kn}^*|, \quad \forall k\in \mathcal K, n\in\mathcal N.
\end{equation}
With the  discrete phase shifts solution $\hat{\boldsymbol \theta}$,  the power control can be obtained
by using the iterative power control scheme  \eqref{Alo3max1dueq5_3}.
According to Fig. \ref{fig311}, it is observed that the performance loss due to the rounding is small for large $b$ and small minimum SINR value $\Gamma$, which indicates that the proposed approach is also suitable to discrete phase shifter with large number of phase shift levels.

\begin{figure}[t]
\centering
\includegraphics[width=3.5in]{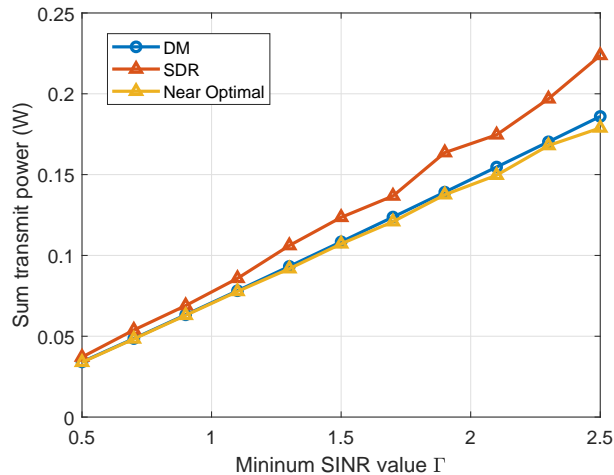}
\vspace{-1em}
\caption{Sum transmit power versus the minimum SINR requirement for DM, SDR, and near optimal solution.} \label{fig312}
\vspace{-1.5em}
\end{figure}

The sum transmit power versus the minimum SINR requirement for DM, SDR, and near optimal solution is given in Fig. \ref{fig312}.
In this figure, the near optimal solution is obtained by the algorithm with two steps. In the first step, the nonconvex unit module constraint is added in the objective function by using the penalty method in \cite{9133435} and then the successive convex approximation method is used to solve the modified optimization problem in the second step. The near optimal solution is calculated by using the successive convex approximation method with multiple initial solutions and the solution with the best objective function is regarded as the near optimal solution.
From Fig. \ref{fig312}, it is shown that the DM always achieves the better performance than that of SDR. The reason is that the SDR scheme requires the  randomizations to construct a rank-one solution, which can lead to the performance degradation.
It can be also seen that DM achieves similar performance with the near optimal solution, which verifies the theoretical findings in Theorem 1.
}

\begin{figure}[t]
\centering
\includegraphics[width=3.5in]{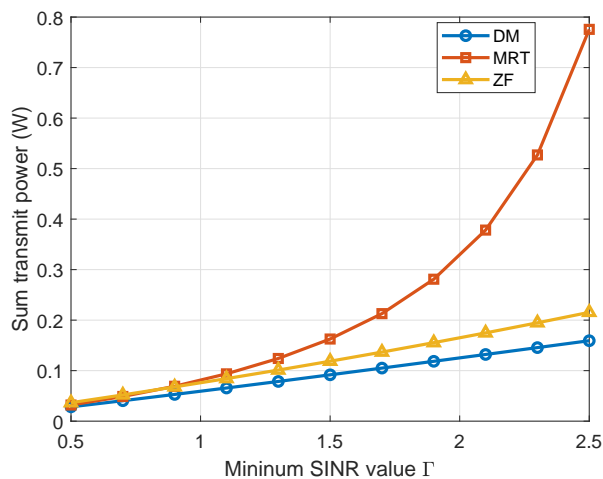}
\vspace{-1em}
\caption{Sum transmit power versus the minimum SINR requirement for DM, MRT, and ZF.} \label{fig2}
\vspace{-1.5em}
\end{figure}

\begin{figure}[t]
\centering
\includegraphics[width=3.5in]{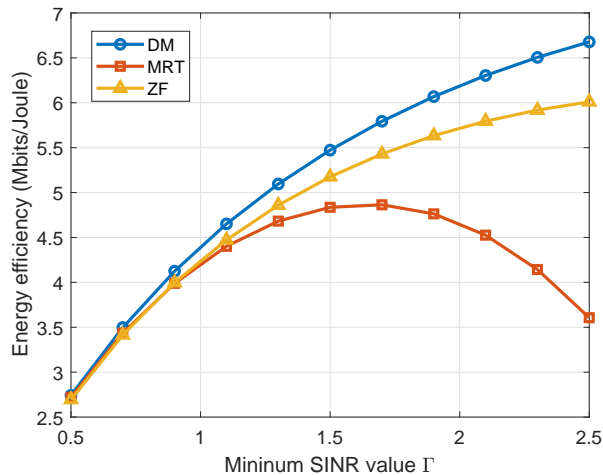}
\vspace{-1em}
\caption{Energy efficiency versus the minimum SINR requirement.} \label{fig3}
\vspace{-1.5em}
\end{figure}

We compare the sum transmit power and energy efficiency performance of DM, MRT, and ZF.
Figs. \ref{fig2} and \ref{fig3} show the sum transmit power and energy efficiency versus the minimum SINR requirement.
{\color{myc1}In Fig. \ref{fig3}, the energy efficiency is calculated by
\begin{equation}
\frac{\sum_{k=1}^KB\log_2(1+\Gamma_k)}{\mu P+{P_{\text{B}}}+{\sum_{k=1}^KP_k}+{NKP_{\text{R}}}},
\end{equation}
where $\mu=\nu^{-1}$ with $\nu=0.8$ being the power amplifier efficiency of the BS,
$P$ is the sum transmit power of the BS,
$P_{\text {B}}=29$ dBm is the circuit power consumption of the BS,
$P_k=5$ dBm is the circuit power consumption of user $k$,
$P_{\text{R}}=5$ dBm is the power consumption of each reflecting element in the RIS,
and $B=1$ MHz is the bandwidth of the BS.}
From these two figures, DM achieves the best performance.
In particular, DM can reduce up to 94\% and 23\% sum transmit power compared to MRT and ZF, respectively.
Besides, DM can increase up to 61\% and 27\% energy efficiency compared to MRT and ZF, respectively.
According to Fig. \ref{fig2}, the sum transmit power increases slightly with the minimum SINR requirement for DM and ZF, while it increases rapidly with the minimum SINR requirement for MRT.
This is due to the fact that MRT only maximizes the received signal strength without considering the multiuser interference, which indicates that MRT is not suitable for high SINR requirement.
It is also found that the energy efficiency of DM and ZF increases with the minimum SINR requirement from Fig. \ref{fig3}.
From both Figs.~\ref{fig2} and \ref{fig3}, it is observed that MRT is superior over ZF for low SINR requirement, while ZF is better than MRT for high SINR requirement.


\begin{figure}[htpb]
\centering
\includegraphics[width=3.5in]{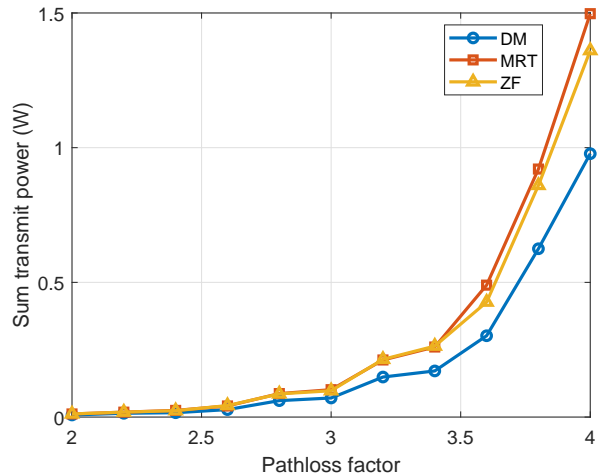}
\vspace{-1em}
\caption{Sum transmit power versus the pathloss factor.} \label{fig32}
\vspace{-1.5em}
\end{figure}

Fig. \ref{fig32} shows how the sum transmit power changes as the pathloss factor varies.
We can see that the sum transmit power of all schemes  increases  with the pathloss factor.
This is because the large pathloss factor results in poor channel gains for the users.
It is found that DM achieves the best performance among all schemes.
From Fig.~\ref{fig32},
DM can achieve the best sum transmit power performance especially for the case that the pathloss factor is high.

\begin{figure}[htpb]
\centering
\includegraphics[width=3.5in]{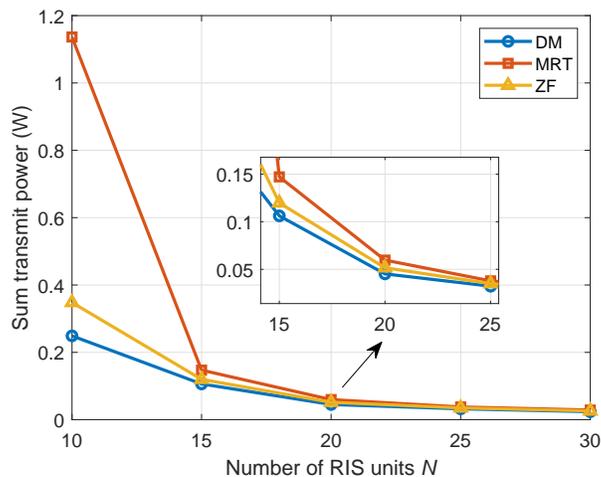}
\vspace{-1em}
\caption{Sum transmit power versus the number of RIS units $N$ with $K=3$.} \label{fig5}
\vspace{-1.5em}
\end{figure}

Fig.~\ref{fig5} shows the sum transmit power versus the number of RIS units $N$.
From this figure, we can see that the sum transmit power of all schemes monotonically decreases with the number of RIS units.
This is because large number of RIS units can lead to high spectral efficiency, which can reduce the transmit power of the system.
According to Fig.~\ref{fig5}, it can be shown that the decrease speed of sum transmit power for MRT is faster than that for ZF and DM, which indicates that MRT is suitable for the case with large number of RIS units.

\begin{figure}[htpb]
\centering
\includegraphics[width=3.5in]{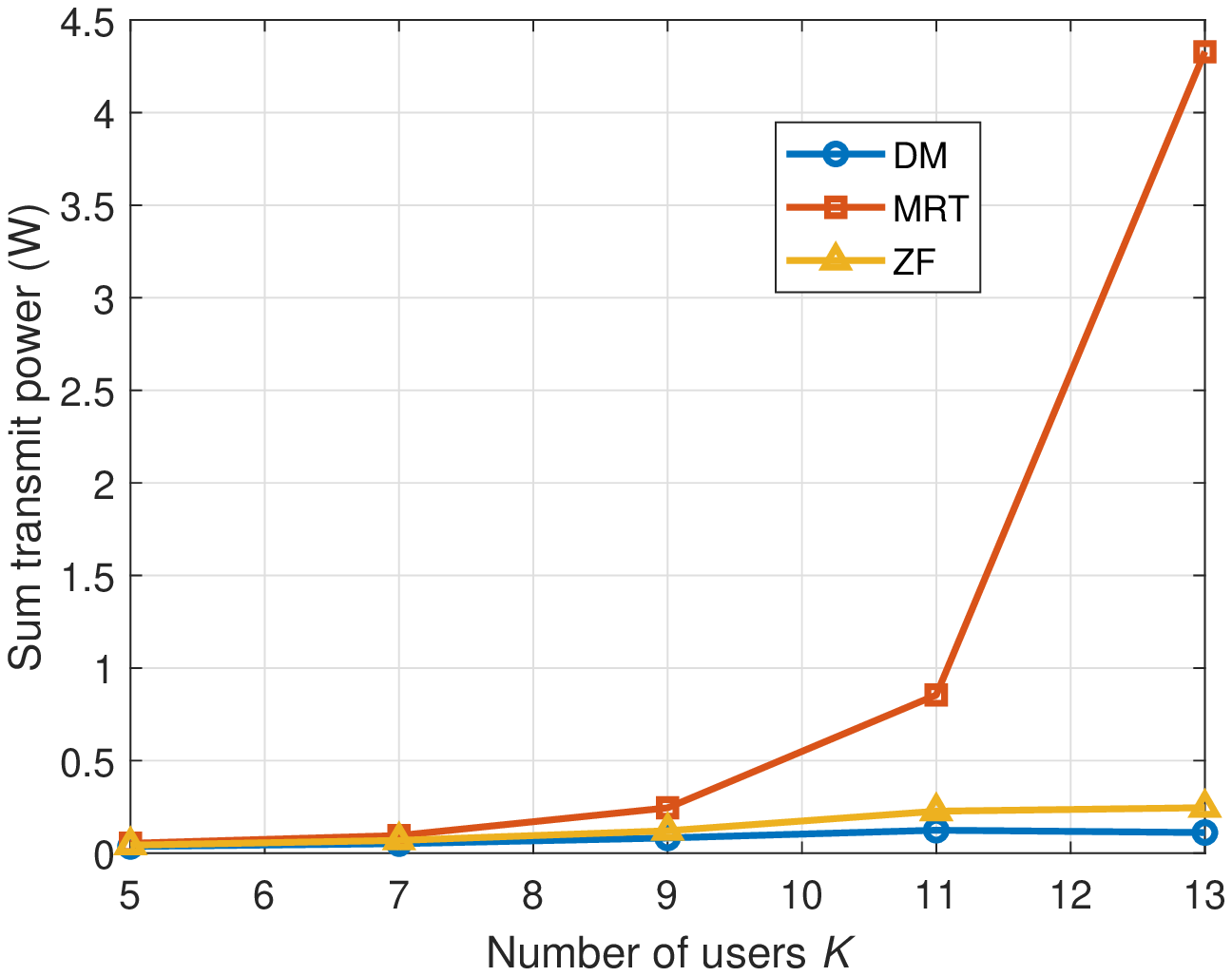}
\vspace{-1em}
\caption{Sum transmit power versus the number of users $K$ with $N=30$.} \label{fig6}
\vspace{-1.5em}
\end{figure}

\begin{figure}[htpb]
\centering
\includegraphics[width=3.5in]{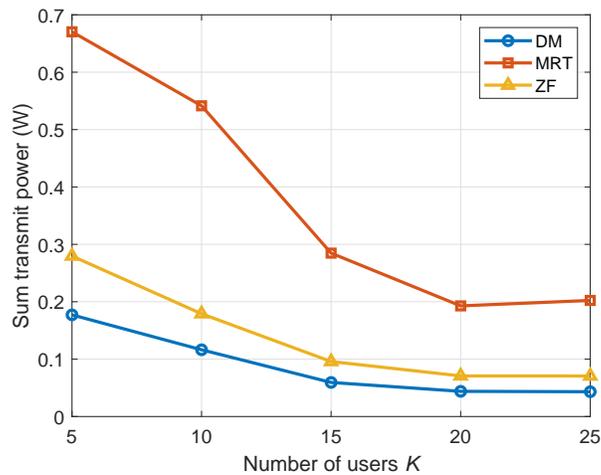}
\vspace{-1em}
\caption{Sum transmit power versus the number of users $K$ with $N=3K$.} \label{fig7}
\vspace{-1.5em}
\end{figure}
Figs.~\ref{fig6} and \ref{fig7} depict the sum transmit power versus number of users with $N=30$ and $N=3K$, respectively.
From both figures, DM achieves the best performance.
In Fig. \ref{fig6}, the sum transmit power increases with the number of users.
This is because the multiuser interference is serious for large number of users.
It can also  be shown that the increase speed of DM is slower than that for MRT or ZF, which shows that
DM can save more power than MRT or ZF.
Clearly, the DM is always better than MRT and ZF especially when the number of users is large.
According to Fig. \ref{fig7}, the sum transmit power tends to decrease with the number of users, which shows the different trends compared to Fig. \ref{fig6}.
The reason is that the number of RIS units is fixed in Fig. \ref{fig6}, while the number of RIS units also increases linearly with the number of users in Fig. \ref{fig7}.
When the ratio between the number of users and the number of RIS units is fixed, the sum transmit power first decreases with the number of users and then remains stable for high number of users.

\begin{figure}[htpb]
\centering
\includegraphics[width=3.5in]{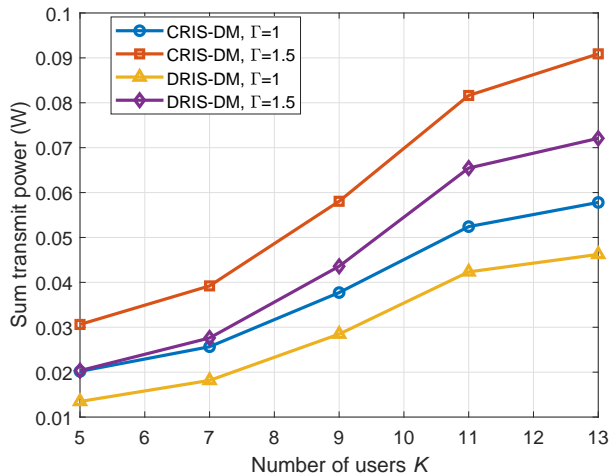}
\vspace{-1em}
\caption{Sum transmit power versus the number of users $K$ for distributed deployment of RIS units with $N=30$.} \label{fig9}
\vspace{-1.5em}
\end{figure}
In Fig. \ref{fig9}, we show the sum transmit power versus the number of users $K$ for distributed deployment of RIS units.
In this figure, the centralized deployment of all RIS units is labeled ``CRIS'', while the distributed deployment of RIS units is labeled ``DRIS''.
In DRIS, we consider the case that there are $K$ RISs and the location of RIS $l$ is given by $(\cos(2l\pi/L), \sin(2l\pi/L))\times100$ m.
For each RIS, it is equipped with $N$ RIS units.
It is found that DRIS achieves the better performance than CRIS, which indicates the benefit of distributed deployment of RISs.
Multiple RISs are spatially distributed in DRIS, which can decrease the transmit power between the transceivers.


%
%
%
%
%


\vspace{-.5em}
\section{Conclusions}\vspace{-.5em}
In this paper, we have investigated the resource allocation problem for a wireless communication network with an RIS-assisted wireless transmitter.
The RIS phase shifts and BS transmit power were jointly optimized to minimize the sum transmit power while
satisfying minimum  SINR requirements and unit-modulus constraints.
To solve this problem, we have proposed the dual method, compared to the SDR, MRT, and ZF beamforming techniques.
Moreover, we have analyzed the asymptotic performance of the RIS-assisted communication system with infinitely number of
RIS units.
Numerical results have shown that the dual method outperforms MRT and ZF schemes in terms of sum transmit power and energy efficiency, especially for high SINR requirements.
Furthermore, the distributed deployment of RIS units was shown to be favorable for decreasing the
sum transmit power.
{\color{myc2}The optimization of RIS elements allocation is left for our future work.}

\bibliographystyle{IEEEtran}
\bibliography{IEEEabrv,ref}

\begin{thebibliography}{10}
\providecommand{\url}[1]{#1}
\csname url@samestyle\endcsname
\providecommand{\newblock}{\relax}
\providecommand{\bibinfo}[2]{#2}
\providecommand{\BIBentrySTDinterwordspacing}{\spaceskip=0pt\relax}
\providecommand{\BIBentryALTinterwordstretchfactor}{4}
\providecommand{\BIBentryALTinterwordspacing}{\spaceskip=\fontdimen2\font plus
\BIBentryALTinterwordstretchfactor\fontdimen3\font minus
  \fontdimen4\font\relax}
\providecommand{\BIBforeignlanguage}[2]{{%
\expandafter\ifx\csname l@#1\endcsname\relax
\typeout{** WARNING: IEEEtran.bst: No hyphenation pattern has been}%
\typeout{** loaded for the language `#1'. Using the pattern for}%
\typeout{** the default language instead.}%
\else
\language=\csname l@#1\endcsname
\fi
#2}}
\providecommand{\BIBdecl}{\relax}
\BIBdecl

\bibitem{saad2019vision}
W.~{Saad}, M.~{Bennis}, and M.~{Chen}, ``A vision of 6g wireless systems:
  Applications, trends, technologies, and open research problems,'' \emph{IEEE
  Netw.}, vol.~34, no.~3, pp. 134--142, 2020.

\bibitem{yu2007transmitter}
W.~Yu and T.~Lan, ``Transmitter optimization for the multi-antenna downlink
  with per-antenna power constraints,'' \emph{IEEE Trans. Signal Process.},
  vol.~55, no.~6, pp. 2646--2660, Jun. 2007.

\bibitem{ngo2013energy}
H.~Q. Ngo, E.~G. Larsson, and T.~L. Marzetta, ``Energy and spectral efficiency
  of very large multiuser {MIMO} systems,'' \emph{IEEE Trans. Commun.},
  vol.~61, no.~4, pp. 1436--1449, Apr. 2013.

\bibitem{buzzi2016survey}
S.~Buzzi, I.~Chih-Lin, T.~E. Klein, H.~V. Poor, C.~Yang, and A.~Zappone, ``A
  survey of energy-efficient techniques for {5G} networks and challenges
  ahead,'' \emph{IEEE J. Sel. Areas Commun.}, vol.~34, no.~4, pp. 697--709,
  Apr. 2016.

\bibitem{zhang2019first}
S.~{Zhang}, S.~{Xu}, G.~Y. {Li}, and E.~{Ayanoglu}, ``First 20 years of green
  radios,'' \emph{IEEE Trans. Green Commun. and Netw.}, vol.~4, no.~1, pp.
  1--15, 2020.

\bibitem{7264975}
W.~{Xu}, Y.~{Cui}, H.~{Zhang}, G.~Y. {Li}, and X.~{You}, ``Robust beamforming
  with partial channel state information for energy efficient networks,''
  \emph{IEEE J. Sel. Areas Commun.}, vol.~33, no.~12, pp. 2920--2935, Dec.
  2015.

\bibitem{huang2020thzris}
C.~Huang, Z.~Yang, G.~C. Alexandropoulos, K.~Xiong, L.~Wei, C.~Chen, and
  Z.~Zhang, ``Hybrid beamforming for ris-empowered multi-hop terahertz
  communications: A {DRL}-based method,'' \emph{Avilable [oneline]:
  https://arxiv.org/abs/2009.09380}, 2020.

\bibitem{pan2019intelligent2}
C.~{Pan}, H.~{Ren}, K.~{Wang}, W.~{Xu}, M.~{Elkashlan}, A.~{Nallanathan}, and
  L.~{Hanzo}, ``Multicell mimo communications relying on intelligent reflecting
  surfaces,'' \emph{IEEE Trans. Wireless Commun.}, vol.~19, no.~8, pp.
  5218--5233, 2020.

\bibitem{pan2019intelligent}
C.~Pan, H.~Ren, K.~Wang, M.~Elkashlan, A.~Nallanathan, J.~Wang, and L.~Hanzo,
  ``Intelligent reflecting surface enhanced {MIMO} broadcasting for
  simultaneous wireless information and power transfer,'' \emph{Avilable
  [oneline]: https://arxiv.org/abs/1908.04863}, 2019.

\bibitem{liu2020reconfigurable}
Y.~Liu, X.~Liu, X.~Mu, T.~Hou, J.~Xu, Z.~Qin, M.~Di~Renzo, and N.~Al-Dhahir,
  ``Reconfigurable intelligent surfaces: Principles and opportunities,''
  \emph{Avilable [oneline]: https://arxiv.org/abs/2007.03435}, 2020.

\bibitem{chongwenDL2020}
C.~{Huang}, R.~{Mo}, and C.~{Yuen}, ``Reconfigurable intelligent surface
  assisted multiuser {{MISO}} systems exploiting deep reinforcement learning,''
  \emph{IEEE J. Sel. Areas Commun.}, vol.~38, no.~8, pp. 1839--1850, Aug. 2020.

\bibitem{huang2019holographic}
C.~{Huang}, S.~{Hu}, G.~C. {Alexandropoulos}, A.~{Zappone}, C.~{Yuen},
  R.~{Zhang}, M.~{Di Renzo}, and M.~{Debbah}, ``Holographic {MIMO} surfaces for
  {6G} wireless networks: {O}pportunities, challenges, and trends,'' \emph{IEEE
  Wireless Commun.}, early access, doi: 10.1109/MWC.001.1900534, 2020.

\bibitem{yang2020energyefficient}
Z.~Yang, M.~Chen, W.~Saad, W.~Xu, M.~Shikh-Bahaei, H.~V. Poor, and S.~Cui,
  ``Energy-efficient wireless communications with distributed reconfigurable
  intelligent surfaces,'' \emph{Avilable [oneline]:
  https://arxiv.org/abs/2005.00269}.

\bibitem{di2020smart}
M.~Di~Renzo, A.~Zappone, M.~Debbah, M.-S. Alouini, C.~Yuen, J.~de~Rosny, and
  S.~Tretyakov, ``Smart radio environments empowered by reconfigurable
  intelligent surfaces: How it works, state of research, and road ahead,''
  \emph{Avilable [oneline]: https://arxiv.org/abs/2004.09352}, 2020.

\bibitem{yang2020RSMARIS}
Z.~Yang, J.~Shi, Z.~Li, M.~Chen, W.~Xu, and M.~Shikh-Bahaei, ``Energy efficient
  rate splitting multiple access ({RSMA}) with reconfigurable intelligent
  surface,'' in \emph{Proc. IEEE Int. Conf. Commun. Workshop}, pp. 1--6, to
  appear, 2020.

\bibitem{long2020reflections}
H.~Long, M.~Chen, Z.~Yang, B.~Wang, Z.~Li, X.~Yun, and M.~Shikh-Bahaei,
  ``Reflections in the sky: Joint trajectory and passive beamforming design for
  secure uav networks with reconfigurable intelligent surface,'' \emph{Avilable
  [oneline]: https://arxiv.org/abs/2005.10559}, 2020.

\bibitem{zhou2020spectral}
S.~Zhou, W.~Xu, K.~Wang, M.~Di~Renzo, and M.-S. Alouini, ``Spectral and energy
  efficiency of {IRS}-assisted {MISO} communication with hardware
  impairments,'' \emph{IEEE Wireless Commun. Lett.}, 2020.

\bibitem{hu2020programmable}
X.~Hu, C.~Zhong, Y.~Zhu, X.~Chen, and Z.~Zhang, ``Programmable metasurface
  based multicast systems: Design and analysis,'' \emph{Avilable [oneline]:
  https://arxiv.org/abs/2002.08611}, 2020.

\bibitem{zhang2020sum}
Y.~Zhang, C.~Zhong, Z.~Zhang, and W.~Lu, ``Sum rate optimization for two way
  communications with intelligent reflecting surface,'' \emph{IEEE Commun.
  Lett.}, vol.~24, no.~5, pp. 1090--1094, 2020.

\bibitem{hum2013reconfigurable}
S.~V. Hum and J.~Perruisseau-Carrier, ``Reconfigurable reflectarrays and array
  lenses for dynamic antenna beam control: {A} review,'' \emph{IEEE Trans.
  Antennas Prop.}, vol.~62, no.~1, pp. 183--198, Jan. 2013.

\bibitem{huang2014relay}
J.~Huang, Q.~Li, Q.~Zhang, G.~Zhang, and J.~Qin, ``Relay beamforming for
  amplify-and-forward multi-antenna relay networks with energy harvesting
  constraint,'' \emph{IEEE Signal Process. Lett.}, vol.~21, no.~4, pp.
  454--458, Apr. 2014.

\bibitem{ntontin2019reconfigurable}
K.~Ntontin, M.~Di~Renzo, J.~Song, F.~Lazarakis, J.~de~Rosny, D.-T. Phan-Huy,
  O.~Simeone, R.~Zhang, M.~Debbah, G.~Lerosey, M.~Fink, S.~Tretyakov, and
  S.~Shamai, ``Reconfigurable intelligent surfaces vs. relaying: {D}ifferences,
  similarities, and performance comparison,'' \emph{Avilable [oneline]:
  https://arxiv.org/abs/1908.08747}, 2019.

\bibitem{jung2019optimality}
M.~Jung, W.~Saad, M.~Debbah, and C.~S. Hong, ``On the optimality of
  reconfigurable intelligent surfaces ({RIS}s): {P}assive beamforming,
  modulation, and resource allocation,'' \emph{Avilable [oneline]:
  https://arxiv.org/abs/1910.00968}, 2019.

\bibitem{yu2020power}
X.~Yu, D.~Xu, D.~W.~K. Ng, and R.~Schober, ``Power-efficient resource
  allocation for multiuser miso systems via intelligent reflecting surfaces,''
  \emph{Avilable [oneline]: https://arxiv.org/abs/2005.06703}, 2020.

\bibitem{xu2020resource}
D.~Xu, X.~Yu, Y.~Sun, D.~W.~K. Ng, and R.~Schober, ``Resource allocation for
  {IRS}-assisted full-duplex cognitive radio systems,'' \emph{Avilable
  [oneline]: https://arxiv.org/abs/2003.07467}, 2020.

\bibitem{8855810}
X.~{Yu}, D.~{Xu}, and R.~{Schober}, ``{MISO} wireless communication systems via
  intelligent reflecting surfaces: (invited paper),'' in \emph{2019 IEEE/CIC
  Int. Conf. Communi. China}, Changchun, China, pp. 735--740, 2019.

\bibitem{hua2020intelligent}
M.~Hua, Q.~Wu, D.~W.~K. Ng, J.~Zhao, and L.~Yang, ``Intelligent reflecting
  surface-aided joint processing coordinated multipoint transmission,''
  \emph{Avilable [oneline]: https://arxiv.org/abs/2003.13909}, 2020.

\bibitem{huang2018achievable}
C.~Huang, A.~Zappone, M.~Debbah, and C.~Yuen, ``Achievable rate maximization by
  passive intelligent mirrors,'' in \emph{Proc. IEEE Int. Conf. Acoust., Speech
  and Signal Process.}, Calgary, Canada, April. 2018, pp. 3714--3718.

\bibitem{jung2018performance}
M.~{Jung}, W.~{Saad}, Y.~{Jang}, G.~{Kong}, and S.~{Choi}, ``Performance
  analysis of large intelligence surfaces ({LIS}s): {A}symptotic data rate and
  channel hardening effects,'' \emph{IEEE Trans. Wireless Commun.}, vol.~19,
  no.~3, pp. 2052--2065, 2020.

\bibitem{zhao2019intelligent}
M.-M. Zhao, Q.~Wu, M.-J. Zhao, and R.~Zhang, ``Intelligent reflecting surface
  enhanced wireless network: {T}wo-timescale beamforming optimization,''
  \emph{Avilable [oneline]: https://arxiv.org/abs/1912.01818}, 2019.

\bibitem{8743496}
H.~{Shen}, W.~{Xu}, S.~{Gong}, Z.~{He}, and C.~{Zhao}, ``Secrecy rate
  maximization for intelligent reflecting surface assisted multi-antenna
  communications,'' \emph{IEEE Commun. Lett.}, vol.~23, no.~9, pp. 1488--1492,
  Sep. 2019.

\bibitem{yu2019robust}
X.~Yu, D.~Xu, Y.~Sun, D.~W.~K. Ng, and R.~Schober, ``Robust and secure wireless
  communications via intelligent reflecting surfaces,'' \emph{Avilable
  [oneline]: https://arxiv.org/abs/1912.01497}, 2019.

\bibitem{9133435}
Q.~{Wu} and R.~{Zhang}, ``Joint active and passive beamforming optimization for
  intelligent reflecting surface assisted {SWIPT} under {QoS} constraints,''
  \emph{IEEE J. Sel. Areas Commun.}, pp. 1--1, 2020.

\bibitem{xu2019resource}
D.~Xu, X.~Yu, Y.~Sun, D.~W.~K. Ng, and R.~Schober, ``Resource allocation for
  secure {IRS}-assisted multiuser {MISO} systems,'' in \emph{Proc. IEEE
  Globecom Workshops (GC Wkshps)}, Waikoloa, HI, USA, pp. 1--6, Dec., 2019.

\bibitem{9133120}
J.~{Liu}, K.~{Xiong}, Y.~{Lu}, D.~W.~K. {Ng}, Z.~{Zhong}, and Z.~{Han},
  ``Energy efficiency in secure {IRS}-aided {SWIPT},'' \emph{IEEE Wireless
  Commun. Lett.}, pp. 1--1, 2020.

\bibitem{guan2020joint}
X.~Guan, Q.~Wu, and R.~Zhang, ``Joint power control and passive beamforming in
  {IRS}-assisted spectrum sharing,'' \emph{IEEE Commun. Lett.}, Jul. 2020.

\bibitem{8741198}
C.~{Huang}, A.~{Zappone}, G.~C. {Alexandropoulos}, M.~{Debbah}, and C.~{Yuen},
  ``Reconfigurable intelligent surfaces for energy efficiency in wireless
  communication,'' \emph{IEEE Trans. Wireless Commun.}, vol.~18, no.~8, pp.
  4157--4170, Aug. 2019.

\bibitem{zhao2018programmable}
J.~Zhao \emph{et~al.}, ``Programmable time-domain digital-coding metasurface
  for non-linear harmonic manipulation and new wireless communication
  systems,'' \emph{National Sci. Rev.}, vol.~6, no.~2, pp. 231--238, 2018.

\bibitem{dai2019wireless}
J.~Y. Dai, W.~K. Tang, J.~Zhao, X.~Li, Q.~Cheng, J.~C. Ke, M.~Z. Chen, S.~Jin,
  and T.~J. Cui, ``Wireless communications through a simplified architecture
  based on time-domain digital coding metasurface,'' \emph{Adv. Mater.
  Technol.}, p. 1900044, 2019.

\bibitem{tang2019programmable}
W.~Tang, J.~Y. Dai, M.~Chen, X.~Li, Q.~Cheng, S.~Jin, K.-K. Wong, and T.~J.
  Cui, ``Programmable metasurface-based {RF} chain-free {8PSK} wireless
  transmitter,'' \emph{Electron. Lett.}, vol.~55, no.~7, pp. 417--420, 2019.

\bibitem{basar2019transmission}
E.~{Basar}, ``Transmission through large intelligent surfaces: A new frontier
  in wireless communications,'' \emph{in Proc. EuCNC}, pp. 112--117, 2019.

\bibitem{basar2019wireless}
E.~{Basar}, M.~{Di Renzo}, J.~{De Rosny}, M.~{Debbah}, M.~{Alouini}, and
  R.~{Zhang}, ``Wireless communications through reconfigurable intelligent
  surfaces,'' \emph{IEEE Access}, vol.~7, pp. 116\,753--116\,773, 2019.

\bibitem{tang2019mimo}
W.~{Tang}, J.~Y. {Dai}, M.~Z. {Chen}, K.~K. {Wong}, X.~{Li}, X.~{Zhao},
  S.~{Jin}, Q.~{Cheng}, and T.~J. {Cui}, ``Mimo transmission through
  reconfigurable intelligent surface: System design, analysis, and
  implementation,'' \emph{IEEE J. Sel. Areas Commun.}, early access, doi:
  10.1109/JSAC.2020.3007055, 2020.

\bibitem{wu2019intelligent}
Q.~Wu and R.~Zhang, ``Intelligent reflecting surface enhanced wireless network
  via joint active and passive beamforming,'' \emph{IEEE Trans. Wireless
  Commun.}, vol.~18, no.~11, pp. 5394--5409, Nov. 2019.

\bibitem{taha2019enabling}
A.~Taha, M.~Alrabeiah, and A.~Alkhateeb, ``Enabling large intelligent surfaces
  with compressive sensing and deep learning,'' \emph{Avilable [oneline]:
  https://arxiv.org/abs/1904.10136}, 2019.

\bibitem{huang2019indoor}
C.~Huang, G.~C. Alexandropoulos, C.~Yuen, and M.~Debbah, ``Indoor signal
  focusing with deep learning designed reconfigurable intelligent surfaces,''
  in \emph{Proc. IEEE 20th International Workshop on Signal Processing Advances
  in Wireless Communications (SPAWC)}, pp. 1--5, 2019.

\bibitem{wei2020parallel}
L.~{Wei}, C.~{Huang}, G.~C. {Alexandropoulos}, C.~{Yuen}, Z.~{Zhang}, and
  M.~{Debbah}, ``Channel estimation for {RIS}-empowered multi-user {MISO}
  wireless communications,'' \emph{Available online:
  https://arxiv.org/abs/2008.01459}, 2020.

\bibitem{1658226}
W.~Yu and R.~{Lui}, ``Dual methods for nonconvex spectrum optimization of
  multicarrier systems,'' \emph{IEEE Trans. Commun.}, vol.~54, no.~7, pp.
  1310--1322, July 2006.

\bibitem{boyd2004convex}
S.~Boyd and L.~Vandenberghe, \emph{Convex optimization}.\hskip 1em plus 0.5em
  minus 0.4em\relax Cambridge University Press, 2004.

\bibitem{grant2008cvx}
M.~Grant, S.~Boyd, and Y.~Ye, ``{CVX: M}atlab software for disciplined convex
  programming,'' 2008.

\bibitem{yang2019eeFL}
Z.~Yang, M.~Chen, W.~Saad, C.~S. Hong, and M.~Shikh-Bahaei, ``Energy efficient
  federated learning over wireless communication networks,'' \emph{Avilable
  [oneline]: https://arxiv.org/abs/1911.02417}, 2019.

\bibitem{yates1995framework}
R.~D. {Yates}, ``A framework for uplink power control in cellular radio
  systems,'' \emph{IEEE J. Sel. Areas Commun.}, vol.~13, no.~7, pp. 1341--1347,
  Sep. 1995.

\bibitem{wu2018intelligent}
Q.~Wu and R.~Zhang, ``Intelligent reflecting surface enhanced wireless network:
  Joint active and passive beamforming design,'' in \emph{Proc. IEEE Global
  Commun. Conf.}, Abu Dhabi, United Arab Emirates, pp. 1--6, Dec. 2018.

\bibitem{so2007approximating}
A.~M.-C. So, J.~Zhang, and Y.~Ye, ``On approximating complex quadratic
  optimization problems via semidefinite programming relaxations,''
  \emph{Mathematical Programming}, vol. 110, no.~1, pp. 93--110, Jun. 2007.

\bibitem{6425521}
S.~{Huang}, H.~{Yin}, J.~{Wu}, and V.~C.~M. {Leung}, ``User selection for
  multiuser {MIMO} downlink with zero-forcing beamforming,'' \emph{IEEE Trans.
  Veh. Technol.}, vol.~62, no.~7, pp. 3084--3097, Sep. 2013.

\bibitem{lobo1998applications}
M.~S. Lobo, L.~Vandenberghe, S.~Boyd, and H.~Lebret, ``Applications of
  second-order cone programming,'' \emph{Linear algebra and its applications},
  vol. 284, no. 1-3, pp. 193--228, 1998.

\bibitem{8846706}
Y.~{Mao}, B.~{Clerckx}, and V.~O.~K. {Li}, ``Rate-splitting for multi-antenna
  non-orthogonal unicast and multicast transmission: {S}pectral and energy
  efficiency analysis,'' \emph{IEEE Trans. Commun.}, vol.~67, no.~12, pp.
  8754--8770, Sept. 2019.

\bibitem{chen2019joint}
M.~Chen, Z.~Yang, W.~Saad, C.~Yin, H.~V. Poor, and S.~Cui, ``A joint learning
  and communications framework for federated learning over wireless networks,''
  \emph{to appear, IEEE Trans. Wireless Commun., Avilable [oneline]:
  https://arxiv.org/abs/1909.07972}, 2020.

\end{thebibliography}

\end{document}